\newif\ifPDFLaTeX
\expandafter\ifx\csname pdfoutput\endcsname\relax\else\PDFLaTeXtrue\fi
\documentclass[12pt,a4paper]{article}
\ifPDFLaTeX
  \usepackage{type1cm}
  \usepackage[pdftex]{feynmp}
  \usepackage[pdftex,colorlinks]{hyperref} 
\else
  \usepackage{feynmp}
\fi
\usepackage{array,amsmath,amssymb}
\usepackage{mathrsfs,url}
\usepackage{thophys}
\allowdisplaybreaks
\newcommand{\thefmfmpsfile}{\thefmffile.\thefmfgraph}
\newcommand{\thefmftexfile}{\thefmffile.t\thefmfgraph}
\makeatletter
\def\fmfgraph(#1,#2){%
  \fmf@graph{#1}{#2}%
  \def\fmfkeep##1{\fmf@keep{#1}{#2}{##1}}%
  \leavevmode
  \IfFileExists{\thefmfmpsfile}%
    {\includegraphics{\thefmfmpsfile}}%
    {\typeout{%
      feynmp: File \thefmfmpsfile\space not found:}}%
  \ignorespaces}
\@namedef{fmfgraph*}(#1,#2){%
  \begin{picture}(#1,#2)
    \fmf@graph{#1}{#2}%
    \def\fmfkeep##1{\fmf@keepstar{#1}{#2}{##1}}%
  \IfFileExists{\thefmfmpsfile}%
    {\put(0,0){\includegraphics{\thefmfmpsfile}}}%
    {\typeout{%
      feynmp: File \thefmfmpsfile\space not found:}}%
      \ignorespaces}
\@namedef{endfmfgraph*}{%
    \endfmfgraph
    \if@fmfio
      {\catcode`\%=14\relax
       \fmf@noexpandoff
        \InputIfFileExists{\thefmftexfile}{}{%
          \typeout{%
       feynmf: Label file \thefmftexfile\space not found:}}}%
    \fi
  \end{picture}}
\makeatother
\renewcommand{\thefmfmpsfile}{\thefmffile\thefmfgraph.eps}
\renewcommand{\thefmftexfile}{\thefmffile\thefmfgraph.tex}
\usepackage{fancyhdr}
\def\preprint#1{\gdef\thepreprint{#1}}
\def\thepreprint{}
\fancypagestyle{fancyplain}{
  \fancyhf{}
  \fancyhead[R]{\thepreprint}
  
  \addtolength{\headsep}{1.0cm}}

\newcommand{\dd}{\mathrm{d}}
\newcommand{\ee}{\mathrm{e}}
\newcommand{\ii}{\mathrm{i}}

\newcommand{\nofrac}[2]{\genfrac{}{}{0pt}{1}{#1}{#2}}
\providecommand{\href}[2]{#2}
\newcommand{\arXiv}[1]{[\href{http://arXiv.org/abs/#1}{#1}]}

\begin{document}
\preprint{WUE-ITP-2003-019(rev)\\MZ-TH/03-17(rev)\\hep-ph/0312263(rev)}
\title{Unitarity, BRST Symmetry and Ward Identities in Orbifold Gauge Theories}
\author{%
  Thorsten Ohl${}^{a}$\thanks{\texttt{ohl@physik.uni-wuerzburg.de}},
  Christian Schwinn${}^{b}$\thanks{\texttt{schwinn@thep.physik.uni-mainz.de}}\\
  \hfil\\
  ${}^{a}$Institut f\"ur Theoretische~Physik und~Astrophysik\\
          Universit\"at~W\"urzburg\\
          Am~Hubland, D-97074~W\"urzburg, Germany\\
  \hfil\\
  ${}^{b}$Institut f\"ur~Physik\\
          Johannes-Gutenberg-Universit\"at~Mainz\\
          Staudingerweg 7,  D-55099 Mainz, Germany}
\date{December 18,~2003\\ (revised June 3,~2004)}
\maketitle
\allowdisplaybreaks
\setlength{\headheight}{42pt}
\thispagestyle{fancyplain}

\begin{abstract}
  We discuss the use of BRST symmetry and the resulting Ward
  identities as consistency checks for orbifold gauge theories in an
  arbitrary number of dimensions.  We demonstrate that both the usual
  orbifold symmetry breaking and the recently proposed Higgsless
  symmetry breaking are consistent with the nilpotency of the BRST
  transformation.  The corresponding Ward identities for
  4-point functions of the theory engender
  relations among the coupling constants that are equivalent
  to the sum rules from tree level unitarity.  We present the complete
  set of these sum rules also for inelastic scattering and discuss
  applications to 6-dimensional models and to incomplete matter
  multiplets on orbifold fixed points.
\end{abstract}

\newpage
\setlength{\headheight}{\baselineskip}
\begin{fmffile}{srpics}
\fmfset{arrow_len}{2mm}

\section{Introduction}
Field theories on a higher dimensional space-time offer new
possibilities for symmetry breaking by orbifold Boundary Conditions
(BCs)
(cf.~e.\,g.~\cite{Antoniadis:1990ew,Quiros:2003gg,Hebecker:2001jb,Hall:2001tn,Barbieri:2001dm})
that have been used to construct novel unified
theories~\cite{Kawamura:1999nj,Hall:2001pg,Altarelli:2001qj,Asaka:2001,Hall:2001xr},
avoiding common problems of 4-dimensional Grand Unified Theories
(GUTs). More general BCs have been used recently in models of
Higgsless Electro-Weak Symmetry
Breaking~(EWSB)~\cite{Csaki:2003dt,Csaki:2003zu,Csaki:fermions}.

Viewed as effective 4-dimensional theories, compactified higher
dimensional field theories are only consistent up to a cutoff usually
associated with the scale of a more fundamental theory, e.\,g.~string
theory.  To make sense of such theories independently of the yet
unknown underlying fundamental theory, a minimal requirement is the
passing of consistency checks like tree level unitarity below the
cutoff and Ward Identities~(WIs).  Tree level unitarity is to be
understood as the requirement that the tree level matrix elements for
$N$-particle scattering amplitudes at high energies scale at most
as~$E^{4-N}$. The more restrictive criterion of partial wave unitarity
shows that $(4+N)$-dimensional gauge theories are valid as effective
theories below the scale of $\sim
g_{D}^{-2/N}$~\cite{SekharChivukula:2001hz,DeCurtis:2002}, where $g_D$
is the dimensionfull $D$-dimensional gauge coupling.

As long as a finite number of fields is involved, tree level unitarity
of theories involving massive gauge bosons requires Spontaneous
Symmetry Breaking~(SSB) by the Higgs
mechanism~\cite{LlewellynSmith:1973,Cornwall:1973}.  This can be
described by a scalar field with coupling constants satisfying
appropriate Higgs sum rules~\cite{Gunion:1991}.  In compactified
higher dimensional theories, the Kaluza-Klein~(KK) modes of the gauge
bosons acquire masses by a geometric Higgs mechanism where the role of
the GBs is played by higher dimensional components of the gauge
bosons.  In 5-dimensional gauge theories, tree level unitarity results
from interlacing cancellations within the infinite gauge boson
KK-tower~\cite{Csaki:2003dt,SekharChivukula:2001hz,DeCurtis:2002,Abe:2003vg}.
These cancellations rely on relations among the coupling constants that
will be called `KK-sum rules' and are helpful in constructing
effective 4-dimensional models for massive gauge bosons without Higgs
bosons~\cite{Csaki:2003dt}.  However, the derivation of these
unitarity Sum Rules~(SRs) for higher dimensional gauge theories from
the explicit calculation of the divergences of the amplitudes is
tedious and examples of SRs have so far only been verified in
specific models~\cite{Hall:2001tn,Abe:2003vg} or have been used
implicitly in the calculation of partial wave unitarity
bounds~\cite{SekharChivukula:2001hz,DeCurtis:2002}.  A first step
toward general SRs has been taken in~\cite{Csaki:2003dt} where two
simple KK-SRs for elastic gauge boson scattering have been derived and
used to check the consistency of Higgsless symmetry breaking.  In
6-dimensional gauge theories that have become popular for $\mathrm{SO}(10)$ GUT
models~\cite{Asaka:2001,Hall:2001xr} and for models of gauge-Higgs
unification~\cite{Csaki:2002ur}, the computation of unitarity
conditions is further complicated by the apparition of physical scalar
components of the gauge fields.  The unitarity of
such models from the KK point of view has not been discussed so far.

Another example for the importance of unitarity as a consistency check
is given by incomplete matter multiplets at the orbifold fixed
points. These are one of the key features of orbifold-GUTs that allow
to implement a natural suppression of Proton decay
and to avoid the doublet-triplet splitting problem.
It has been checked for the example of boundary Higgs bosons in
a $\mathrm{SU}(5)$ theory that this explicit symmetry breaking doesn't cause
unitarity violations~\cite{Hall:2001tn}, but a more general discussion
of the consistency of this setup has not been given.

However, tree level unitarity by itself is not sufficient for the
consistency of the theory. In theories of vector bosons, invariance of
the gauge fixed action under a nilpotent BRST transformation is
crucial for a consistent quantization with a unitary
$S$-matrix~\cite{Kugo:1979}.  Therefore the nilpotency of the BRST
transformation and checks of BRST invariance of the scattering
amplitudes---i.\,e.~checks of the appropriate WIs---are important
criteria for the consistency of a theory besides the verification of
tree level unitarity.

According to~\cite{LlewellynSmith:1973,Cornwall:1973}, gauge
invariance and tree level unitarity are equivalent in Spontaneously
Broken Gauge Theories~(SBGTs) with renormalizable couplings.  To use
the WIs as a tool for consistency checks, one should, however,
determine a minimal set of amplitudes that has to be checked to ensure
the consistency of the theory on tree level. 
In this paper, we determine a set of WIs that
allows for simple, model independent and comprehensive gauge
checks~\cite{Schwinn:2003}.

This result allows us to give a much simpler derivation of the
conditions following from the unitarity SRs.  The WIs are easier to
implement both in model building and in checking numerical amplitudes
for phenomenological calculations.

In section~\ref{sec:ed-gauge} we introduce the KK-decomposition of the
gauge boson lagrangian in an arbitrary number of dimensions and
perform the gauge fixing. We verify that both the usual orbifold BCs
and the Dirichlet BCs employed in Higgsless EWSB are consistent with
the nilpotency of the BRST transformation and allow to
define physical states and derive WIs similar to a 4-dimensional
SBGT.
In section~\ref{sec:higgs-sr} we discuss the use of tree level
unitarity and WIs as consistency checks.  We review the SRs derived
from unitarity and demonstrate that they can be equivalently obtained
by imposing simple WIs on the 4-point scattering matrix elements.  In
section~\ref{sec:kk-sr} we apply the SRs to unitarity cancellations in
6-dimensional gauge theories, demonstrating the important role of the
physical scalar components of the gauge bosons which are a new feature
compared to 5-dimensional gauge theories.  In section~\ref{sec:brane}
we apply our results to analyze the consistency of incomplete matter
multiplets on orbifold fixed points.

\section{BRST-symmetry and Ward Identities in orbifold gauge theories}
\label{sec:ed-gauge}

To derive WIs for a KK-gauge theory that are similar to those of a
4-dimensional SBGTs, one has to choose an appropriate gauge fixing and
introduce a nilpotent BRST transformation that leaves the gauge fixed
KK-lagrangian invariant. In KK-theories on orbifolds, care has to be
taken to use consistent BCs at the orbifold fixed points.  In previous
discussions~\cite{Hebecker:2001jb,Csaki:2003dt}, consistency with the
equations of motion and unitarity has been used as a criterion. Here
we will introduce the nilpotency of the BRST transformation as a
practical criterion that is much easier to use than the
calculation of unitarity violating terms of scattering amplitudes
in~\cite{Csaki:2003dt}.

\subsection{Kaluza Klein decomposition of a $Z_2^n$ orbifold gauge theory}
\label{sec:kk}

To establish our notation and to
introduce an appropriate gauge fixing, we perform the KK-decomposition
of a $Z_2$ orbifold gauge theory in $4+N$ dimensions.  Subsequently we
will derive the WIs and determine BCs consistent with the nilpotency
of the BRST transformation.

For definiteness, we assume a factorisable constant metric of the
$(4+N)$-dimensional space-time of the form $
g_{AB}=\text{diag}(\eta_{\mu\nu}, -\gamma_{ij})$.  The generalization
to a warped background metric by plugging appropriate warp factors
into the formulae below is straightforward. 
Our notation for the Yang-Mills lagrangian in $4+N$ dimensions is 
given in appendix~\ref{app:kk-lag}.

The KK-decomposition of the gauge fields is
\begin{equation}
\label{eq:kk-decomp}
  A^a_A(x,y)=
      \begin{pmatrix}
        A^a_\mu(x,y)\\
        \Phi^a_i(x,y)
      \end{pmatrix}
  = \sum_{\vec n}
      \begin{pmatrix}
        f^a_{\vec n}(y) A^a_{\vec n,\mu}(x)\\
        g^a_{\vec n}(y) \Phi^a_{\vec n,i}(x)
      \end{pmatrix}
\end{equation}
where the wavefunctions $\chi=f,g$ satisfy the differential equation
\begin{equation}
\label{eq:kk-dgl}
  \partial_i\partial^i \chi_{\vec n}^a(y)=-{m^a_{\vec n}}^2\chi^a_{\vec n}
\end{equation}
They are chosen orthonormal and satisfying a completeness relation:
\begin{equation}
\label{eq:complete}
  \int\!\dd^N y\, \chi^a_{\vec n}(y) \chi^a_{\vec m}(y)=\delta_{\vec n,\vec m}
  \,,\,
   \sum_{\vec n} \chi^a_{\vec n}(x)\chi^a_{\vec n}(y)=\delta(y-x)
\end{equation}
(the group indices~$a$ are not summed over here).  To determine the BCs
imposed on the wavefunctions, let us recall some aspects of symmetry
breaking on
orbifolds~\cite{Quiros:2003gg,Hebecker:2001jb,Hall:2001tn,Barbieri:2001dm}.
An orbifold $C/K$ is obtained from a compact manifold $C$ and a
discrete group $K$ by identifying points $y\in C$ under the action of
$K$, i.\,e.~$y \simeq P_ky$, where the $P_k$ form a representation of
$K$ on $C$.  In an orbifold the action of $K$ has a set of fixed
points $\{y_f\}$
\begin{equation}
  y_{f}=P_k y_{f}
\end{equation}
for some $k\in K$. (For more than one extra dimension, there can also
be fixed lines, fixed surfaces etc.).  Fields defined on an orbifold
need only be invariant under $K$ up to transformations $Z_k$ of a
symmetry group of the lagrangian that form a representation of $K$ in
field space:
\begin{equation}
\Phi\left(P_ky\right)=Z_k \Phi(y)
\end{equation}
For definiteness, we will discuss the most familiar example of an
orbifold symmetry: the group $K=Z_2$ that is generated by one element
$P(y-y_f) = -(y-y_f)$. In this case\footnote{%
  For more complicated orbifold symmetries like $T^2/Z_4$ in
  6~dimensions~\cite{Csaki:2002ur}, the orbifold transformation can
  mix higher dimensional components, since a homogeneous
  transformation of the covariant derivative requires the
  transformation law
  $\Phi^a_i(x,P_ky)=\eta^a_{k}\Phi^a_j(x,y)(P_k^{-1})_{ji}$.
  Therefore the BCs of the $\Phi_i$ depend on the matrix $P_k$ and the
  general description of the KK-decomposition becomes more involved.}
the transformations of the gauge fields take the form
\begin{equation}
\label{eq:orbi-trans}
  \begin{aligned}
    A^a_\mu(x,P y ) &=  \eta^aA^a_\mu(x,y)\\
    \Phi^a_i(x,P y) &= -\eta^a\Phi^a_i(x,y)
  \end{aligned}
\end{equation}
where the $\eta^a$ are the eigenvalues of the representation
matrix~$Z$.  The transformation law of the higher dimensional
components is determined from the requirement of a homogeneous
transformation of the covariant derivative.  For~$\eta^a\neq 1$,
the 4-dimensional gauge fields must vanish at the fixed
points~$y_{f}$, i.\,e.~they must satisfy Dirichlet BCs. Therefore the gauge
symmetry  is broken to a subgroup $H_k$ at the boundary. 
The different parity of the scalar components implies that they
satisfy Neumann BCs.  For~$\eta^a= 1$ the symmetry remains
unbroken and the BCs  of 4-dimensional vectors and scalars are
exchanged.  This can be summarized as (identifying the indices
corresponding to broken generators by a hat):
\begin{equation}
\label{eq:orbi-bcs}
  \begin{aligned}
          A^{\hat a}_\mu(x,y_f)&=0 \qquad &\partial_i \Phi^{\hat a}_j(x,y_f)&=0\\
    \partial_i A^{a}_\mu(x,y_f)&=0 \qquad &\Phi^a_i(x,y_f)                  &=0\\
  \end{aligned}
\end{equation}
Only gauge fields that remain unbroken at every fixed point have
KK-zero-modes with vanishing masses. If desired, zero-modes of the
higher dimensional scalar components of the broken gauge fields that
survive the orbifolding~\eqref{eq:orbi-trans} can be projected out by
introducing further orbifold symmetries.

The conditions~\eqref{eq:orbi-bcs} translate into the
boundary conditions for the KK-wavefunctions 
\begin{equation}
  \begin{aligned} 
    f^{\hat a}(y_f)        &=0 \qquad &\partial_i g^{\hat a}(y_f) &=0\\
    \partial_i f^{ a}(y_f) &=0 \qquad &                g^{ a}(y_f)&=0
  \end{aligned}
\end{equation}
Because they satisfy the same BCs, the derivatives $\partial_i f$ can
be expanded in the basis of the~$g$ and the~$\partial_i g$ can be
expanded in the basis of the~$f$.  One can 
choose the~$g$ such  that
\begin{equation}
\label{eq:wf-relation}
  \partial_i f_{\vec n} = m_{n_i} g_{\vec n}\quad,\quad 
  \partial_i g_{\vec n}= -m_{n_i} f_{\vec n}
\end{equation}
with $\sum_i m_{n_i}^2=m_{\vec n}^2$.
This is consistent with the equation of motion~\eqref{eq:kk-dgl} and
will diagonalize the gauge boson masses and coupling between the
$A_\mu$ and the $\Phi_i$.  For the familiar torus compactification on
orbifolds, the property~\eqref{eq:wf-relation} is obviously satisfied
and a similar relation has been imposed for the warped case
in~\cite{Randall:2001gb}. 

The effective 4-dimensional lagrangian can now be derived using the
KK-decomposition~\eqref{eq:kk-decomp} and exploiting the
relations~\eqref{eq:wf-relation}.  To avoid notational clutter, we
introduce multi-indices $\alpha\equiv(a,\vec n)$ and
$\alpha_i\equiv(a,0,\ldots,0,n_i,0,\ldots,0)$ and use a summation
convention also for the sum over the KK-states.  Using the
relations~\eqref{eq:wf-relation}, we find the cubic interaction terms
\begin{multline}
\label{eq:kk-int}
  \mathscr{L}^{KK}_{\text{cubic}} =
    - g^{\alpha\beta\gamma}\partial_\mu A_\nu^\alpha A^{\beta,\mu}A^{\gamma,\nu} 
    - \frac{1}{2}T^{\alpha}_{\beta\gamma}\, A^{\alpha,\mu}
            \Phi^{\beta,i}\overleftrightarrow{\partial_\mu} \Phi_i^\gamma\\
    + \frac{1}{2}g_{\Phi AA}^{i\alpha\beta\gamma}
            \Phi^{\alpha,i} A_{\mu}^{\beta}A^{\mu,\gamma}
    - \frac{1}{2} T^{\alpha}_{\beta\gamma}
            (m_{\alpha_j}\Phi^\alpha_{i}-m_{\alpha_i}\Phi^\alpha_{j})
            \Phi^{\beta,i}\Phi^{\gamma,j}\,.
\end{multline}
The remaining terms in the KK-lagrangian are given in~\eqref{eq:kk-lag}  
in the appendix. The coupling
constants are products of group theory factors and integrals over
products of KK-wavefunctions, e.\,g.:
\begin{subequations}
\label{eq:kk-couplings}
\begin{align}
\label{eq:g_abc}
  g^{\alpha \beta\gamma}
    &= f^{abc}\int\!\dd^N y\, f^\alpha (y)f^\beta(y) f^\gamma (y)\\
\label{eq:t_abc}
  T^{\alpha}_{\beta\gamma}
    &= f^{abc}\int\!\dd^N y \, f^\alpha (y)g^\beta (y)g^\gamma (y)\,.
\end{align}
\end{subequations}
The explicit form of the other couplings is given 
in~\eqref{eq:kk-couplings-app}.  Note that
the interactions among the scalars originate from the~$F_{ij}$
components and appear only in more than one extra dimension.

The KK-decomposition yields a bilinear mixing $-m_{\alpha_i}\partial_\mu
\Phi^{\alpha,i} A^{\alpha,\mu}$ of 4-di\-mensional gauge bosons and
scalars. Therefore at each KK-level the linear combination
\begin{equation}
\label{eq:KK-gb}
 \phi^\alpha\equiv -\frac{m_{\alpha_i}}{m_\alpha}\Phi^{\alpha,i}
\end{equation}
plays the role of a geometric Goldstone boson that is eaten by the KK
gauge bosons, leaving $N-1$ physical scalars. The sign
in~\eqref{eq:KK-gb} is chosen because of compatibility with our
conventions for the WIs in 4-dimensional theories.

Accordingly, we decompose  the scalars into 
the GBs~\eqref{eq:KK-gb} and `geometric Higgs' bosons:
\begin{equation}
\label{eq:scalar-decompose}
  \Phi^{\alpha}_i = H^{\alpha}_i - \frac{m_{\alpha_i}}{m_{\alpha}}\phi^{\alpha}
\end{equation}
where $H^i$ is defined as orthogonal to the GBs,
i.\,e.~$m_{\alpha_i}H^{\alpha,i}=0$.  The mass term for the scalar
obtained from the KK-decomposition of the $F_{ij}^2$ term of the
lagrangian has the form
\begin{equation}
  m_{\alpha}^2 {\Phi^{\alpha,i}}^2 -(m_{\alpha_i}\Phi^{\alpha,i})^2=
  m_{\alpha}^2 {H^{\alpha,i}}^2
\end{equation}
and the GBs are massless, as they must be.  To eliminate the mixing of
gauge bosons and GBs, we choose a gauge fixing function
\begin{equation}
\label{eq:gf}
  G^a=-\frac{1}{\xi}(\partial_\mu A^{a,\mu}(x,y)-\xi \partial_i\Phi^{a,i}(x,y))
\end{equation}
that extends the one introduced
in~\cite{Mueck:2001,SekharChivukula:2001hz} to more than one extra
dimension.  In terms of KK-modes, the gauge fixing lagrangian takes
the form
\begin{equation}
\label{eq:gauge-fix}
  \mathscr{L}_{GF}=-\int\!\dd^N y\,
  \frac{1}{2\xi}{G^a}^2=-\frac{1}{2\xi}
        (\partial_\mu A^{\alpha,\mu}-\xi m_{\alpha}\phi^{\alpha})^2
\end{equation}

\subsection{BRST-Symmetry and consistent boundary conditions}
\label{sec:brs}
The possible symmetry breaking patterns resulting from orbifold
BCs~\eqref{eq:orbi-trans} are highly
constrained~\cite{Hebecker:2001jb}.  For example, neither EWSB nor the
breaking of $\mathrm{SO}(10)$ to the SM is possible in 5~dimensions by
abelian orbifold conditions alone.  These constraints arise from the
relation
\begin{equation}
\label{eq:auto}
  f^{abc}=\eta^a_{k} \eta^b_{k} \eta^c_{k}f^{abc} 
\end{equation}
following from the requirement that the field strength transforms
according to $F^a_{\mu\nu}\to \eta^a_{k} F^a_{\mu\nu}$. Because
of~\eqref{eq:auto}, only structure constants with an even number of
broken indices can be nonvanishing.

In order to liberate model builders from these constraints, mixed BCs
of the form
\begin{equation}
\label{eq:mixed}
  \partial_i A_\mu^{a}(y_f)=V^{ab}_{y_f}A_\mu^b(y_f)
\end{equation}
have been proposed in~\cite{Hebecker:2001jb}.  They can be introduced
consistently by coupling the gauge fields to a Higgs boson at the
boundary~\cite{Hebecker:2001jb,Csaki:2003dt}. However,
imposing such BCs without including the Higgs boson engenders
unitarity violations.  Taking the vacuum expectation value of the
boundary Higgs to infinity, the mixed BCs turn into Dirichlet BCs
\begin{equation}
\label{eq:dirichlet}
  A_\mu^b(y_f)=0
\end{equation}
that maintain the unitarity of gauge boson scattering, while avoiding
the constraints from orbifold symmetry breaking. This possibility has
been utilized for Higgsless EWSB by BCs
alone~\cite{Csaki:2003dt,Csaki:2003zu,Csaki:fermions}.

To investigate the mixed and Dirichlet BCs further, we will use the
nilpotency of the BRST quantization as a consistency check.  
To derive the BRST transformations of the KK-modes we use that
consistency of the
$(4+N)$-dimensional BRST transformation of the gauge boson
\begin{equation}
  \delta_{\text{BRST}} A^a_{M}(x,y)=\partial_M c^a(x,y)
     + f^{abc}A_{M}^b(x,y) 
       c^c(x,y)
\end{equation}
demands that ghosts must satisfy the same BCs as the
4-dimensional components of the gauge bosons and therefore have a
KK-decomposition in terms of the $f^\alpha$. 
From the BRST
transformation of the antighost together
with the equation of motion of the auxiliary field, 
it can be inferred that the same
wavefunctions appear also in the KK decomposition of the antighosts.  
The complete set of BRST transformations in $(4+N)$-dimensions is given
 in~\eqref{eq:d-brs}. 
The BRST
transformations of the KK-modes are then given by
\begin{subequations}
\label{eq:brs}
  \begin{align}
    \delta_{\text{BRST}} A^\alpha_{\mu}(x)&=\partial_\mu c^\alpha(x)
     + g^{\alpha\beta\gamma}A_{\mu}^\beta(x) 
       c^\gamma(x)\\
    \delta_{\text{BRST}} \Phi^\alpha_{i}(x)&=m_{\alpha_i} c^\alpha(x)
     +T^\gamma_{\alpha\beta}\Phi_{i}^\beta(x) 
       c^\gamma(x)\label{eq:brs-gold}\\
    \delta_{\text{BRST}} c^\alpha(x) &=
        -\frac{1}{2}g^{\alpha\beta\gamma}c^\beta(x)c^\gamma(x)\\
    \delta_{\text{BRST}} \bar c^\alpha(x)&=B^\alpha(x)\\
    \delta_{\text{BRST}} B^\alpha(x)&=0
  \end{align}
\end{subequations}
and the gauge fixing~\eqref{eq:gf} implies the equations of motion
of the KK-modes of the auxiliary field $B$:
  \begin{equation}\label{eq:b-eom}
  B^\alpha = -\frac{1}{\xi}G^\alpha
     = -\frac{1}{\xi}(\partial_\mu A^{\alpha,\mu}-\xi m_{\alpha}\phi^{\alpha})
\end{equation}
The appearance of
the inhomogeneous term in the BRST transformations~\eqref{eq:brs-gold}
of the higher dimensional components of the gauge fields supports
the interpretation of~\eqref{eq:KK-gb} as GBs. 
The remaining transformation laws 
agree with those of a gauge theory with (an infinite number of)
`structure constants' $g^{\alpha\beta\gamma}$ and `generators'
$T^{\gamma}_{\alpha\beta}$.

The BRST transformations are nilpotent iff the relations
\begin{subequations}
\label{eq:BRS-cond}
\begin{align}
\label{eq:KK-lie}
  T^{\gamma}_{\alpha\beta}T^{\delta}_{\beta\epsilon}-T^{\delta}_{\alpha\beta}
   T^\gamma_{\beta\epsilon}&=
  g^{\beta\delta\gamma}T^{\beta}_{\alpha\epsilon} \\
\label{eq:KK-cons-lie}
  m_{\beta_i}T^{\gamma}_{\alpha\beta}
  -m_{\gamma_i}T^{\beta}_{\alpha\gamma}&=m_{\alpha_i} g^{\alpha\beta\gamma}\\
\label{eq:KK-jacobi}
  g^{\alpha\beta\epsilon}g^{\gamma\delta\epsilon}+g^{\gamma\alpha\epsilon}
  g^{\beta\delta\epsilon}+g^{\alpha\delta\epsilon}g^{\beta\gamma\epsilon}&=0
\end{align}
\end{subequations}
hold. Inserting the definitions~\eqref{eq:g_abc} and~\eqref{eq:t_abc}
and using the relations~\eqref{eq:wf-relation} we find that the
condition~\eqref{eq:KK-cons-lie} (a similar relation occurs in a
4-dimensional SBGT, cf.~\eqref{eq:2phi-w-coupling}) is equivalent to
\begin{equation} 
  0=f^{abc}\int\!\dd^Ny\, \partial_i (g^\alpha f^\beta f^\gamma)=
      f^{abc}\lbrack g^\alpha f^\beta f^\gamma \rbrack_{y_f}
\end{equation}
For three unbroken indices, the boundary term vanishes, because
$g^\alpha$ is zero on the boundary.  For one broken index, the
structure constants vanish, because the unbroken generators must close
into an algebra.  For two or three broken indices, there is at least
one broken wavefunction $f^{\hat \alpha}$. Thus~\eqref{eq:KK-cons-lie} 
is satisfied as long as the broken wavefunctions vanish on the boundary.
This shows that general
Dirichlet BCs~\eqref{eq:dirichlet} are consistent, in contrast to the
mixed BCs~\eqref{eq:mixed}.  This confirms the results obtained from
unitarity in~\cite{Csaki:2003dt}.

The `Lie Algebra'~\eqref{eq:KK-lie} and the `Jacobi
Identity'~\eqref{eq:KK-jacobi} are satisfied automatically by the
Jacobi identity of the structure constants $f^{abc}$.
This can be seen using the completeness relations of the
KK-wavefunctions (cf.~also~\cite{Csaki:2003dt}).  Inserting the
definitions of the coupling constants~\eqref{eq:kk-couplings} and
suppressing the group indices for the moment, every term
of~\eqref{eq:KK-lie} and~\eqref{eq:KK-jacobi} involves an expression
of the form
\begin{multline}
  \sum_{\vec n}
    \int\!\dd^Ny\,    \chi_{\vec n}(y) \chi_{\vec m}(y) \chi_{\vec l}(y)
    \int\!\dd^N{y'}\, \chi_{\vec n}(y')\chi_{\vec k}(y')\chi_{\vec p}(y')\\
  = \int\!\dd^Ny\,    \chi_{\vec m}(y) \chi_{\vec l}(y) \chi_{\vec k}(y) \chi_{\vec p}(y)
\end{multline}
Here we have used the completeness relations~\eqref{eq:complete} to get
rid of the sum over the KK-modes.
As an example, the RHS of~\eqref{eq:KK-lie} can be put in the form
\begin{equation}
\label{eq:complete-simplify}
  g^{\beta\delta\gamma}T^{\beta}_{\alpha\epsilon}
    = f^{bdc}f^{bae}\int\!\dd^Ny\, f^{\delta}(y) f^\gamma(y)g^\alpha (y)g^\epsilon(y)
\end{equation}
Performing similar manipulations for the remaining terms, the same
integral over the KK-wavefunctions appears everywhere
and~\eqref{eq:KK-lie} is reduced to the Jacobi Identity for the
$f^{abc}$.  The identity~\eqref{eq:KK-jacobi} can be treated
accordingly.

Having ensured the nilpotency of the BRST-charge, we can perform the
gauge fixing as usual by adding the BRST transform of a functional of
ghost number $-1$:
\begin{equation}
\label{eq:fp}
  \mathscr{L}_{GF}+\mathscr{L}_{FP}=\delta_{\text{BRST}}
    \left[\bar c_\alpha(G^\alpha+\frac{\xi}{2}B^\alpha)\right]
\end{equation}
with the gauge fixing functional $G^\alpha$ as defined
in~\eqref{eq:gf}.  Having a BRST invariant gauge fixed lagrangian and
a nilpotent BRST charge, the physical states can be defined as usual
by the Kugo-Ojima condition
\begin{equation}
  Q \ket{\psi_{\text{phys}}}=0 
\end{equation}
Using the equation of motion for the auxiliary field
$B^a$~\eqref{eq:b-eom}, we obtain the WI
\begin{equation}
  0=\braket{\phi_{\text{phys}}|\{Q, \bar c^\alpha\}|\psi_{\text{phys}}}
   =-\frac{1}{\xi}\braket{\phi_{\text{phys}}|
       (\partial_\mu A^{\alpha,\mu}-\xi m_{\alpha}\phi^{\alpha}) |\psi_{\text{phys}}}
\end{equation}
To turn this into an identity for scattering matrix elements, one needs
to amputate the external gauge boson and GB propagator. With our
choice of gauge fixing, the usual $R_\xi$ gauge tree level relation of
the propagators~\cite{Chanowitz:1985}
\begin{equation}
\label{eq:prop-rel}
  k_\mu D^{\mu\nu}_A=-\xi D_\phi k^\nu
\end{equation}
is satisfied\footnote{%
  In higher orders, loop corrections to~\eqref{eq:prop-rel} have to be
  considered similar to the 4-dimensional case~\cite{Yao:1988}.}
and the amputation results in the WI for scattering matrix elements
\begin{equation}
\label{eq:gf-wi}
  -\ii p_{a\mu} \mathcal{M}^\mu(A^{\alpha}(p_a)\dots)
       -m_\alpha\mathcal{M}(\phi^\alpha(p_a)\dots)
  \equiv \mathcal{M}(\mathcal{D}_a(p_a)\dots) = 0
\end{equation}
that is similar to the WI in a 4D SBGT. This will allow us in
section~\ref{sec:higgs-sr} to use the SRs obtained from the WIs in
4-dimensional theories also for KK-theories.

\section{Unitarity Sum Rules and Ward Identities}
\label{sec:higgs-sr}
The WIs~\eqref{eq:gf-wi} implied by BRST symmetry provide a powerful tool 
for consistency checks in gauge theories.
An advantage over tree unitarity as consistency criterion is the 
simpler applicability due to the fact that the WIs hold at every point
in phase space, independent of the external momenta. In contrast, 
tree level unitarity requires to take the high energy limit of the
momenta. Therefore WIs are also more powerful as consistency checks
in numerical calculations~\cite{Schwinn:2003}. 

As we demonstrate in this section, the SRs of the coupling constants
implied by tree level unitarity, along with additional relations for
GB couplings, can also be obtained from a suitable set of WIs for
4-point functions. Our discussion is valid for general field theories
with the field content of a SBGT and dimension four couplings,
including orbifold gauge theories.
We establish our result by computing the WIs using a general parametrisation
of a lagrangian for such a theory  without assuming the symmetry relations
resulting from a spontaneously broken gauge symmetry.  A comparison with
the unitarity SRs~\cite{LlewellynSmith:1973,Cornwall:1973,Gunion:1991}
then shows that both approaches yield the same results.  The
applicability of the SRs to KK-gauge theories is discussed in
subsection~\ref{sec:kk-wis}.

The key formula derived below and used in the discussion of unitarity in 6-dimensional gauge
theories in section~\ref{sec:kk-sr} is the SR~\eqref{eq:gold-lie}. 
The discussion
of incomplete matter multiplets on the boundary in section~\ref{sec:brane}
makes use of the Lie algebra structure of the matter gauge 
couplings~\eqref{eq:fermion-lie}.
Apart from serving as consistency checks, the SRs discussed in this section
provide also tools in model building. In the context of non minimal Higgs
models, this has been discussed already in~\cite{Gunion:1991} while for
models of EWSB without a Higgs, a simple SR obtained from elastic 
scattering has been used in~\cite{Csaki:2003dt}. As a possible future
application of the SRs, it would be interesting to use the SRs to
explore 4-dimensional UV completions of the Higgsless models
 of~\cite{Csaki:2003dt,Csaki:2003zu}, for instance by truncating 
the KK tower and introducing heavy
scalars with approaprite couplings determined by the SRs
to cancel unitarity violations in the scattering of 
the higher KK-levels. 
\subsection{A minimal set of Ward Identities}
To use the WIs~\eqref{eq:gf-wi} of SBGTs
as a tool for consistency checks, one should determine a minimal set
of amplitudes that has to be checked to ensure the consistency of the
theory.  Since the Feynman rules of SBGTs are determined by tree level
unitarity of 4 particle scattering amplitudes~\cite{Cornwall:1973}
(apart from the scalar self interaction that make it necessary to
consider some 5-point amplitudes), it is reasonable to expect that a
similar set of amplitudes is sufficient for the consistency checks
using the WIs. However, this connection has never be made precise in the 
literature.

In an unbroken gauge theory, it is not difficult to see that the
Yang-Mills structure of the lagrangian can indeed be `reconstructed'
by imposing the WI on the 4-point amplitudes. It is a well known
textbook example, that the WI of the two quark-two gluon amplitude
implies the Lie algebra relations of the quark-gluon couplings.
Similarly one can show that the WI of the 4-gluon amplitude determines
the quartic gluon coupling and implies the Jacobi Identity for the
triple gluon coupling. As discussed in subsection~\ref{sec:wi-srs}, 
the same relations are also obtained in SBGTs.

However, in SBGTs in $R_\xi$ gauge,
the determination of the \emph{complete} set of Feynman rules
from WIs with one contraction is more involved, since 
amplitudes with external GBs have to be considered. 
Because of the unphysical nature of the GBs, such WIs 
include ghost terms in addition to~\eqref{eq:gf-wi} and further
consistency checks would be required to determine the ghost Feynman
rules.  As we will show below, such complications 
can be avoided by considering the generalized WIs
\begin{equation}
\label{eq:chanowitz}
  \mathcal{M}(\mathcal{D}\dots\mathcal{D}\Phi\dots\Phi)=0
\end{equation}
with up to 4 contractions that allow to reconstruct the Feynman rules
of SBGTs (apart from quartic Higgs selfcouplings) from the
\emph{limited} set of WIs for the 4-point functions without external
GBs.  This allows for simple, model independent and comprehensive
gauge checks that avoid the introduction of ghosts already on tree
level.

Since the lagrangian of a KK-gauge theory has the same form 
as that of a SBGT and the same WIs~\eqref{eq:gf-wi} hold,
we can then infere that the same SRs are also valid
for a KK-theory. However, the need to truncate potentially
divergenent sums over KK-modes might spoil the unitarity cancellations, 
this is discussed in subsection~\ref{sec:kk-wis}.
\subsection{Unitarity sum rules from Ward Identities}
\label{sec:wi-srs}
To present the results of the computation of the WIs, we have to introduce
a parametrization of the general renormalizable Lagrangian  with the
particle spectrum of a SBGT but \emph{without} imposing gauge
invariance.  We denote gauge bosons by $W^a$, Goldstone bosons by
$\phi_a$ and all other scalar fields by $H_i$. We will not introduce a
separate notation for massless gauge bosons and take it as understood
that the GBs are only associated to massive gauge bosons.
The couplings that will be most relevant below are defined as
 \begin{multline}\label{eq:general-int}
    \mathscr{L}_{\text{int}}
      = - f^{abc}W_{b,\mu} W_{c,\nu}\partial^\mu W_a^\nu
        + \frac{1}{2}g_{\phi WW}^{Aab}\phi_A W_{a,\mu}W_b^\mu
        - \frac{1}{2}T^a_{AB}(\phi_A\overleftrightarrow{\partial_\mu}\phi_B)W_a^\mu\\
      + \bar\psi_i\fmslash W_a\left[
        \tau^{a}_{Lij}(\tfrac{1-\gamma^5}{2})+
        \tau^{a}_{Rij}(\tfrac{1+\gamma^5}{2})\right]\psi_j
       + \bar\psi_i \phi_A \left[X_{ij}^A (\tfrac{1-\gamma^5}{2})
                        +{X_{ij}^{A}}^\dagger (\tfrac{1+\gamma^5}{2})\right]\psi_j 
\end{multline}
Here we have collected GBs and physical scalars in a vector
$   \phi_A=(\phi_a, H_i)$.
The components of the coupling matrices of the scalars will be denoted as
\begin{equation}
\label{eq:t-def}
  \mathbf{T}^c=
    \begin{pmatrix}
      t^c &  -g_{H\phi W}^{c}\\ (g_{H\phi W}^c)^T & T^c
    \end{pmatrix}
  \quad,\quad
  \mathbf{X}=
    \begin{pmatrix}
      g_\phi^a\\ g_{H}^k
    \end{pmatrix}
\end{equation}
The complete lagrangian in this notation
is given in appendix~\ref{app:general-fr}. The 
translation to the notation used in the specific example of the KK-theory 
in~\eqref{eq:kk-int} should be clear.

We use the lagrangian~\eqref{eq:general-int} without further assumptions in 
the evaluation of the STIs. In order to describe a SBGT, however,
the couplings appearing in~\eqref{eq:general-int}
must satisfy certain invariance conditions. Apart from the familiar 
Jacobi Identity of the $f^{abc}$ and the Lie algebra of the fermion gauge 
couplings 
\begin{subequations}
\label{subeq:unitarity-conditions}
\begin{equation}
\label{eq:fermion-lie}
[\tau_{L/R}^a, \tau_{L/R}^b]_{ij}=\ii f^{abc}\tau_{L/Rij}^c\\
\end{equation}
also the gauge couplings of the scalars must satisfy a Lie algebra
and the Yukawa couplings must be invariant under global transformations
generated by the $\tau$ and $T$:
\begin{align}
\label{eq:scalar-lie}
  \lbrack\mathbf{T}^a,\mathbf{T}^b\rbrack&=f^{abc}\mathbf{T}^c\\
\label{eq:yukawa-tensor}
  -\ii \tau^{a}_R\mathbf{X}+i\mathbf{X}\tau^a_L &=\mathbf{T}^a\mathbf{X}
\end{align}
\end{subequations}
There are also constraints on the quartic couplings, for instance
the quartic couplings of the scalars to gauge boson in a SBGT have the form
\begin{equation}
\label{eq:2s2w}
  g_{\phi^2 W^2}^{ABcd}=-\{{\bf T}^c,{\bf T}^d\}_{AB}
\end{equation}
Furthermore,
the condition that some of the scalars $H_i$ aquire a vacuum
expectation value that is responsible for gauge boson and fermion masses
allows to express the couplings $g_{\phi WW}$ in terms of the ${\bf T}$
and the masses
and furthermore to eliminate all the GB couplings as independent
parameters (cf.~appendix~\ref{app:gb-couplings}).
With these identifications, the relations~\eqref{subeq:unitarity-conditions}
and the component $g_{H^2 W^2}$ of~\eqref{eq:2s2w}
 are precisely the unitarity SRs derived  
in~\cite{LlewellynSmith:1973,Cornwall:1973,Gunion:1991}. 

We now
demonstrate how the unitarity SRs~\eqref{subeq:unitarity-conditions} and~\eqref{eq:2s2w}  can be obtained also from WIs of four point functions. 
Since this requires the somewhat tedious task to expand the SRs in component
form and compare to the results of the WIs, we have collected the
explicit expressions in appendix~\ref{app:wi-results}. Details on the calculation
of the WIs can be found in~\cite{Schwinn:2003}.

In a first step, we evaluate the WIs for the 3-point matrix elements
and find they determine the form of the cubic couplings involving GBs,
reproducing the expressions in appendix~\ref{app:gb-couplings}.
As examples, the results of these WIs 
determine the GB components of the matrices~\eqref{eq:t-def}
\begin{subequations}
\label{subeq:unitarity-relations}
\begin{align}
\label{eq:2phi-w-coupling}
   t_{ac}^b&=-\frac{1}{2m_{W_a}m_{W_c}}f^{abc}(m_{W_b}^2-m_{W_a}^2-m_{W_c}^2)\\
\label{eq:hphiw-coupling}
   m_{W_a} g_{H\phi W}^{iab}&=-\frac{1}{2}g_{HWW}^{iab}\\
\label{eq:f-phi-coupling}
   m_{W_a} g_{\phi ij}^a &= \ii(m_j \tau^{a}_{Rij}- m_i\tau^{a}_{Lij})
\end{align}
\end{subequations}
While the same definitions appear in the unitarity approach as auxiliary
quantities~\cite{LlewellynSmith:1973}, the physical content
of~\eqref{subeq:unitarity-relations} remains, however, rather obscure in
this approach because they are not identified with GB couplings. 
The remaining relations obtained from the 3 point WIs are given 
in appendix~\ref{app:reconstruct_3point} 
and include conditions that are a consequence of the gauge invariance of the
scalar potential and
haven't been derived from unitarity in~\cite{LlewellynSmith:1973}.

We now turn to the evaluation of WIs for 4-point scattering
amplitudes with one contraction.
The results can be found in appendix~\ref{app:physical} 
and include  consistency relations among the
coupling constants of the physical particles like the Lie algebra
structure of the fermion couplings~\eqref{eq:fermion-lie},
the Higgs component of the Lie algebra~\eqref{eq:scalar-lie}, the
invariance condition of the Higgs-Yukawa coupling contained
in~\eqref{eq:yukawa-tensor} or the expression for the $WWHH$
coupling in~\eqref{eq:2s2w}.
As an example, the WI for the 4 gauge boson amplitude 
implies the Jacobi Identity of the $f^{abc}$ and determines the
quartic gauge boson coupling:
\begin{equation}
  \label{eq:4w-wi}
  \parbox{20mm}{%
    \begin{fmfchar}(20,20)
      \fmfleft{A1,A2}
      \fmfright{A3,A4}
      \fmf{double}{A1,a}
      \fmf{photon}{A2,a}
      \fmf{photon}{A3,a}
      \fmf{photon}{A4,a}
      \fmfv{decor.shape=circle,decor.filled=empty,decor.size=25}{a}
    \end{fmfchar}}\Rightarrow 
    \begin{cases}
      f^{abe}f^{cde}+f^{cae}f^{bde}+f^{ade}f^{bce}=0\\
      g_{W^4}^{abcd}=f^{abe}f^{cde}-f^{ade}f^{bce}
    \end{cases}
\end{equation}
In the diagrammatical representation of the WIs, the insertion of 
the operator $(\partial_\mu W^\mu-m_W\phi)$ is represented by a double line.

As a general rule, the results obtained from the WIs with one contraction
correspond to the SRs ensuring the cancellation of the leading divergencies 
in the unitarity approach~\cite{LlewellynSmith:1973,Cornwall:1973,Gunion:1991}.
The relations~\eqref{eq:4w-wi}, for instance arise from the cancellation
of the terms growing like $E^4$ in gauge boson scattering.

Turning to the results of the WIs with more than one contraction, 
it turns out that they include the remaining components of the Lie
algebra~\eqref{eq:scalar-lie} and the Yukawa transformation
law~\eqref{eq:yukawa-tensor}. These are the
SRs ensuring the cancellation of the
subleading divergences in the unitarity approach so
indeed all the SRs ensuring tree level unitarity
can be obtained in a simpler way from the WIs.
As an example, let us give the relation that ensures the cancellation
of the subleading divergencies~$\propto E^2$ in gauge boson 
scattering~\cite{LlewellynSmith:1973,Cornwall:1973,Gunion:1991} 
that we will verify in a 6-dimensional KK-theory in section~\ref{sec:kk-sr}.
In our approach, it arises from the WI for the 4-gauge boson amplitude
with three contractions and is given by 
\begin{equation}
  \label{eq:gold-lie}
  \parbox{20mm}{%
    \begin{fmfchar}(20,20)
      \fmfleft{A1,A2}
      \fmfright{A3,A4}
      \fmf{double}{A1,a}
      \fmf{double}{A2,a}
      \fmf{double}{A3,a}
      \fmf{photon}{A4,a}
      \fmfv{decor.shape=circle,decor.filled=empty,decor.size=25}{a}
    \end{fmfchar}}\Rightarrow 
  \begin{aligned}
    \Biggl\{ \frac{1}{m_{W_e}^2}
       \left[f^{abe}f^{ced}(m_{W_a}^2-m_{W_b}^2-m_{W_e}^2)
             (m_{W_c}^2-m_{W_e}^2-m_{W_d}^2)\right]\\
      - g_{HWW}^{iab}g_{HWW}^{icd}\Biggr\} - \quad a\leftrightarrow c 
         =  2f^{ace}f^{ebd}(m_{W_e}^2-m_{W_b}^2-m_{W_d}^2)
  \end{aligned}
\end{equation}
We take it as understood that the internal summations involving inverse
gauge boson masses extend only over massive gauge bosons.
The resulting equation for elastic scattering has been 
used in~\cite{Csaki:2003dt} to check the consistency of EWSB by
Dirichlet boundary conditions in 5 dimensions.
Using the result~\eqref{eq:2phi-w-coupling} for the GB-gauge boson coupling, 
one can check that~\eqref{eq:gold-lie} is the $ab$ component of the
Lie-algebra of the scalar couplings ~\eqref{eq:scalar-lie}.

Apart from these relations ensuring unitarity of gauge boson scattering,
the results of the remaining WIs of 4-point functions 
given in appendix~\ref{app:wi-results}  include further conditions 
determining the GB couplings that have not been obtained from
unitarity in~\cite{LlewellynSmith:1973,Cornwall:1973,Gunion:1991}.
At one hand, these are  the conditions
\begin{equation}
\label{eq:2s-jac}
  T^a_{AB}g^{Bbc}_{\phi WW}-m_{W_a}g^{Aabc}_{\phi^2 W^2}
   =f^{acd}g^{Abd}_{\phi WW}+f^{abd}g^{Acd}_{\phi WW}
\end{equation}
that fix the quartic GB-gauge boson couplings.
This relations follow from the graded Jacobi Identity
\begin{equation}
\label{eq:super-jacobi}
  0= \lbrack A,\{B,C\} \rbrack + \{\lbrack C, A \rbrack,B\} - \{ \lbrack A,B\rbrack C\}
\end{equation}
and the form of the 2-gauge boson 2-scalar
interaction~\eqref{eq:2s2w}.  

The remaining WIs of 4 point functions imply the condition for the gauge invariance of the scalar 
potential:
\begin{equation}
\label{eq:quartic-scalar}
   m_{W_\alpha}g^{ABCa}_{\phi^4}
      +g^{DBC}_{\phi^3}T^a_{DA}
      +g^{ADC}_{\phi^3}T^a_{DB}
      +g^{ABD}_{\phi^3}T^a_{DC}=0
\end{equation}
This relation determines the quartic couplings of the GBs in terms of
already known quantities. 
Apart from the quartic Higgs selfcouplings $g_{\phi^4}^{ijkl}$ that are not 
included in~\eqref{eq:quartic-scalar}, the WIs of four point
functions therfore determine the relations among the coupling constants of
a SBGT and fix the GB couplings in terms of the couplings of 
the physical particles.

\subsection{Sum rules in KK-theories}
\label{sec:kk-wis}
The KK-decomposition of a higher dimensional gauge theory results in
an effective 4-dimensional lagrangian of the same general
form~\eqref{subeq:ssb-lag}, albeit with an infinite number of fields.
As discussed in subsection~\ref{sec:brs}, the gauge fixing
function~\eqref{eq:gf} and a consistent set of BCs ensures the
validity of WIs similar to those of a 4-dimensional SBGT and the same
SRs also hold in KK-theories.  However, in calculations of scattering
amplitudes, sums over KK-modes appear that can be infinite if
KK-momentum conservation breaks down.  For more than one extra
dimension, the sums over the propagators diverge and a cutoff has to
be introduced. Therefore the connection of the SRs to unitarity
requires further clarification for KK-theories.

Let us first note that independently of the convergence properties of
the amplitudes, no divergencies appear in the unitarity SRs, even
though they involve---like the conditions ~\eqref{eq:BRS-cond}
considered in subsection~\ref{sec:brs}---potentially infinite sums
over products of `generators' and coupling constants
from~\eqref{eq:kk-couplings}.  However, just as
in~\eqref{eq:complete-simplify} we can use the completeness
relations~\eqref{eq:complete} and the sum over the KK-modes drops out
of all SRs and no problems with divergences arise. 
To infer that the SRs obtained from
the 4-point WIs also imply tree level unitarity for the scattering
amplitudes of a KK-gauge theory, we have to use, however, that the sum
over the KK-modes converges, since a regularization procedure can lead
to a violation of the SRs.  For a sharp cutoff of the sum, the SRs
cannot be satisfied for the highest KK-level~\cite{Csaki:2003dt}.  The
introduction of a smooth regulator in the KK-sum corresponds to a
modification of the KK-propagators.  Because the gauge (and unitarity)
cancellations rely on WIs connecting contracted vertices and inverse
propagators, they can be spoiled by such a modification.

We expect however, that this poses no serious problem in realistic
applications: As in 4-dimensional SBGTs, the unitarity SRs doesn't
imply partial wave unitarity. Instead of the usual 4-dimensional bound
on the Higgs mass $\sim 1$ TeV~\cite{Lee:1977}, in the 5-dimensional
SM there is an upper bound of the number of KK-modes $N_{KK}\sim
10$~\cite{SekharChivukula:2001hz} compatible with unitarity.  Since
this bound is rather small compared the scale where problems from
diverging KK-sums arise, partial wave unitarity is expected to give a
more stringent criterion for the breakdown of unitarity than the
disturbance of unitarity cancellations by a cutoff.

In general, since problems with unitarity are associated with
longitudinally polarized massive gauge bosons, a conflict of unitarity
and regularization can arise only in models where the couplings of
gauge bosons violate KK-selection rules in such a way that an infinite
number of modes contributes to scattering processes with external
gauge bosons.  Furthermore there must be more than one extra dimension
allowing the KK-sums to diverge.  The models considered in
sections~\ref{sec:kk-sr} and \ref{sec:brane} are not affected by such
problems.  However, these issues deserve further studies, especially
because one expects that possible violations of gauge invariance by a
cutoff will play a more important role in loop calculations.

\section{Unitarity Sum Rules in 6 dimensions}
\label{sec:kk-sr}
As an application of the unitarity SRs, we will discuss gauge boson
scattering in a 6-dimensional gauge theory.  The new ingredients
compared to the 5-dimensional theories discussed
in~\cite{Csaki:2003dt,SekharChivukula:2001hz,DeCurtis:2002,Abe:2003vg}
are the physical scalar component of the bulk gauge bosons
(cf.~\eqref{eq:scalar-decompose}) and the apparition of more than one
KK-state at the same level. As we will demonstrate, these two
phenomena are connected and the physical scalars play an important
role in the unitarity cancellations if the initial and final state
KK-momenta are not parallel. 

While the SRs considered in this work only ensure the correct scaling
behavior implied by unitarity, we expect our results also to be useful
as a first step in a partial wave analysis along the lines
of~\cite{SekharChivukula:2001hz,DeCurtis:2002}.  Other interesting
possible extensions of our work include a compactification with
different radii and nontrivial shape parameters~\cite{Dienes:2001wu},
the introduction of additional orbifold fixed points for the
application to GUT models~\cite{Asaka:2001,Hall:2001xr} and more
complicated orbifold symmetries like $Z_4$~\cite{Csaki:2002ur}.

\subsection{Kaluza-Klein decomposition and Feynman rules}
\label{sec:6d}

For simplicity, we consider a 6-dimensional gauge theory
compactified on a torus with identical radii without symmetry breaking
by orbifold BCs.  Before orbifolding, 
the KK-decomposition of an arbitrary field $\Phi$ is
given by
\begin{equation}
  \Phi(x,y)=\frac{1}{2\pi R}\sum_{\vec n}\Phi_{\vec n}(x)
  \ee^{\frac{\ii}{R}(\vec n\cdot \vec y)}
\end{equation}
where $\vec n=(n_1,n_2)$.  The KK-modes satisfy
$\Phi_{\vec n}^*=\Phi_{-\vec n}$ for hermitian fields.

According to~\eqref{eq:KK-gb}, the unphysical GB and the orthogonal
physical component are~\cite{Asaka:2001}
\begin{subequations}
\label{eq:6d-scalars}
\begin{align}
  \phi^a_{\vec n}&=-\frac{1}{R m_{\vec n}}\left(n_1\Phi^{a,1}_{\vec n}+n_2
    \Phi^{a,2}_{\vec n}\right)\\
  H^a_{\vec n}&=\frac{1}{R m_{\vec n}}
    \left(-n_2\Phi_{\vec n}^{a,1}+n_1\Phi_{\vec n}^{a,2}\right)
\end{align}
\end{subequations}

We will limit our discussion to a simple
orbifolding~\cite{Appelquist:2000nn} $T^2/Z_2$ with $Z_2: (y,z)\to
(-y,-z)$ where all gauge symmetries remain unbroken. This imposes the
conditions $\Phi^+_{n1,n2}=\Phi^+_{-n1,-n2}$ on the KK-modes with
positive $Z_2$ parity, while the negative parity modes must satisfy
$\Phi^-_{n1,n2}=-\Phi^-_{-n1,-n2}$.  The range of the coordinates in
the orbifold can be taken as~\cite{Appelquist:2000nn} $-\pi R\leq
y_1\leq \pi R$ and $0\leq y_2 \leq \pi R$.

The decompositions of the vector and scalar fields become:
\begin{align}
  A^{a,\mu}(x,y)=&\frac{1}{\sqrt{2}\pi R}A^{a,\mu}_{0}(x)
     +\frac{1}{\pi R}\sum_{\vec n}^\infty
         \cos\left(\frac{ n_iy_i}{R}\right)A^{a,\mu}_{\vec n}(x)\\
  \Phi^{a,i}(x,y)=&\frac{1}{\pi R^2}\sum_{\vec n }
     \frac{1}{m_{\vec n}}\sin\left(-\frac{ n_iy_i}{R}\right)\Phi^{a,i}_{\vec n}(x)\\
\label{eq:phi-decomp}
  \Phi^{a,i}_{\vec n}(x)&=\left[-\binom{n_1}{n_2}\phi^{a}_{\vec n}(x)
     +\binom{-n_2}{n_1} H^{a}_{\vec n}(x) \right]
\end{align}
with the summation range\footnote{%
  This differs slightly from the
  conventions in~\cite{Appelquist:2000nn} where the condition
  $n_1+n_2\geq 1\;\text{or}\;n_1=-n_2\geq 1$ is used. Both conventions
  lead to the same degeneracies at each KK-level, e.\,g.~our level
  $(2,\pm 1)$, $(1,\pm 2)$ corresponds to $(2,\pm 1)$, $(\pm 1,2)$ in
  the conventions of~\cite{Appelquist:2000nn}. For general
  considerations our convention seems to be more practical because we
  don't have to distinguish $|n_1|<|m_1|$ from $|n_1|>|n_2|$.}
\begin{equation}
   n_1\geq 1 \text{ and } |n_2| \geq 1 \;\text{or}\; n_i=0 \text{ and } n_j\geq 1
\end{equation}
The sign conventions in~\eqref{eq:phi-decomp} have been chosen in
accordance with~\eqref{eq:wf-relation} and~\eqref{eq:KK-gb}.
Symmetry breaking can be introduced by imposing further $Z_2$
symmetries with fixed points $y,z=\frac{1}{2}\pi R$~\cite{Asaka:2001}.

The interactions of the KK-modes  involve KK-number conservation 
factors of the form (following the notation of~\cite{Mueck:2001})
\begin{equation}
\begin{aligned}
  \delta_{\vec k,\vec l,\vec m}&=
     \delta_{\vec k+\vec l+\vec m,0}
    +\delta_{\vec k+\vec l-\vec m,0}
    +\delta_{\vec k-\vec l+\vec m,0}
    +\delta_{\vec k-\vec l-\vec m,0}\\
  \tilde\delta_{\vec k,\vec l,\vec m}&=
    -\delta_{\vec k+\vec l+\vec m,0}
    +\delta_{\vec k+\vec l-\vec m,0}
    -\delta_{\vec k-\vec l+\vec m,0}
    +\delta_{\vec k-\vec l-\vec m,0}
\end{aligned}
\end{equation}
The triple gauge boson interactions and the scalar coupling have the
same form as in the 5-dimensional case~\cite{Mueck:2001}:
\begin{align}
\label{eq:6d-3g}
  g^{\alpha\beta\gamma}&=
     f^{abc} \left(\frac{1}{\sqrt 2}\right)^{(1+\delta_{\vec n,0}
      +\delta_{\vec m,0}+\delta_{\vec k,0})}\delta_{\vec n,\vec m,\vec k}\\
\label{eq:6d-a2phi}
  g_{\Phi AA}^{i\alpha\beta\gamma}&=
    gf^{abc} \left[\frac{1}{\sqrt 2^{\delta_{\vec k,0}+1}}
       \tilde\delta_{\vec n,\vec m,\vec k}\frac{m_i}{R}
         - \frac{1}{\sqrt 2^{\delta_{\vec m,0}+1}}
              \tilde\delta_{\vec n,\vec k,\vec m}\frac{k_i}{R}\right]
\end{align}
where we have introduced the 4-dimensional coupling $g=g_6/(\sqrt 2\pi R)$.

Inserting the decomposition of $\Phi$ in terms of Higgs and GBs, we
find for the interaction of the Higgses
\begin{multline}
\label{eq:higgs-coupling}
  g_{HAA}^{\alpha\beta\gamma}=
    gf^{abc} \Biggl[\frac{1}{\sqrt 2^{\delta_{\vec k,0}+1}}
    \tilde\delta_{\vec n,\vec m,\vec k}\frac{-m_1 n_2+m_2n_1}{m_{\vec n}R^2}\\
  -\frac{1}{\sqrt 2^{\delta_{\vec m,0}+1}}
    \tilde\delta_{\vec n,\vec k,\vec m}\frac{-k_1 n_2+k_2n_1}{m_{\vec n}R^2}\Biggr]
\end{multline}

\subsection{Sum rules for elastic scattering}
We are now ready to investigate the SRs for gauge boson scattering.
We begin with the simplest case of scattering at the same KK-level
where the initial and final KK-states $\vec n$ are the same. Because
of the KK-number conservation the contributing diagrams are
\begin{equation}
\parbox{30mm}{%
  \fmfframe(5,5)(5,5){%
    \begin{fmfchar*}(20,20)
      \fmftop{A1,A2} \fmfbottom{A3,A4} 
      \fmf{photon}{A1,a} 
      \fmf{photon}{A2,b}
      \fmf{photon,label=$\nofrac{0}{2\vec n}$}{a,b} 
      \fmf{photon}{A3,a} 
      \fmf{photon}{A4,b}
      \fmfdot{a}
      \fmfdot{b}  
      \fmfv{label=$\vec n$}{A1,A2,A3,A4}
    \end{fmfchar*}}}\qquad
\parbox{30mm}{%
  \fmfframe(5,5)(5,5){%
    \begin{fmfchar*}(20,20)
      \fmftop{A1,A2} \fmfbottom{A3,A4} \fmf{photon}{A1,a} \fmf{photon}{A2,a}
      \fmf{photon,label=$\nofrac{0}{2\vec n}$}{a,b} 
      \fmf{photon}{A3,b} 
      \fmf{photon}{A4,b}
      \fmfdot{a}
      \fmfdot{b}  
    \fmfv{label=$\vec n$}{A1,A2,A3,A4}
    \end{fmfchar*}}}\qquad
\parbox{30mm}{%
  \fmfframe(5,5)(5,5){%
    \begin{fmfchar*}(20,20)
      \fmftop{A1,A2} \fmfbottom{A3,A4} 
      \fmf{photon}{A1,a} 
      \fmf{photon}{A3,b} 
      \fmf{photon,label=$\nofrac{0}{2\vec n}$}{a,b} 
      \fmf{phantom}{A4,b}
      \fmf{phantom}{A2,a}
      \fmffreeze
      \fmf{photon}{A4,a}
      \fmf{photon}{A2,b}
      \fmfdot{a}
      \fmfdot{b}  
      \fmfv{label=$\vec n$}{A1,A2,A3,A4}
    \end{fmfchar*}}}\quad
\parbox{25mm}{%
  \fmfframe(5,5)(5,5){%
    \begin{fmfchar*}(15,15)
      \fmftop{A1,A2} \fmfbottom{A3,A4} 
      \fmf{photon}{A1,a} 
      \fmf{photon}{A3,a} 
      \fmf{photon}{A4,a}
      \fmf{photon}{A2,a}
      \fmfdot{a}
      \fmfv{label=$n$,la.di=5}{A1,A2,A3,A4}
    \end{fmfchar*}}}
\end{equation}
From the explicit form of the Higgs
couplings~\eqref{eq:higgs-coupling} we see that the Higgs couplings
vanish for the scattering at the same KK-level.  The
SR~\eqref{eq:gold-lie} obtained equivalently from the
$\mathcal{M}(\mathcal{D}\mathcal{D}\mathcal{D}W)$ WI or the
cancellation of the subleading divergences in quartic gauge boson
scattering then reduces to the relation discussed already
in~\cite{Csaki:2003dt}
\begin{equation}
\label{eq:4n-sr}
   3m_{\vec m}^2g^{ace}_{\vec n,\vec n,\vec m}g^{ebd}_{\vec m,\vec n,\vec n}
  =4m_{\vec n}^2g^{ace}_{\vec n,\vec n,\vec m}g^{bde}_{\vec n,\vec n,\vec m}
\end{equation}
where we have used the Jacobi Identity.  Inserting the coupling
constants~\eqref{eq:6d-3g} (note that $\delta_{\vec n,\vec
m,0}=2\delta_{\vec n,\vec m}$)
\begin{equation}
\label{eq:6d-3w}
  g^{abc}_{\vec n,\vec n,0} = g f^{abc}\,,\;
  g^{abc}_{\vec n,\vec m,\vec n+\vec m} = \frac{g}{\sqrt 2} f^{abc}
\end{equation}
we find that this is indeed satisfied in the same way as in the
5-dimensional case discussed in~\cite{Abe:2003vg}.

A new phenomenon in theories with more than one extra dimension is
scattering at the same KK-level, but with different KK-numbers at the
initial and final state, e.\,g.~KK-exchange
\begin{equation}
  \vec n= \binom{n_1}{n_2}\quad \to \quad\vec m=\binom{n_2}{n_1}
\end{equation}
Here, because the second component of the KK-quantum numbers can
take negative values, the exchange of 
vector bosons with the 
quantum numbers
\begin{equation}
  \vec k_+=\binom{n_1+n_2}{n_1+n_2} \quad
  \vec k_-=\binom{|n_1-n_2|}{-|n_1-n_2|}\quad
   m_{\vec k_{\pm}}=\sqrt 2 \frac{|n_1\pm n_2|}{R}
\end{equation}
is possible in the $t$-and $u$-channel:
\begin{subequations}\label{eq:4n-diag}
\begin{equation}
\parbox{30mm}{\fmfframe(5,5)(5,5){%
  \begin{fmfchar*}(20,20)
    \fmftop{A1,A2} \fmfbottom{A3,A4} 
    \fmf{photon}{A1,a} 
    \fmf{photon}{A2,b}
    \fmf{photon,label=0}{a,b} 
    \fmf{photon}{A3,a} 
    \fmf{photon}{A4,b}
    \fmfdot{a}
    \fmfdot{b}  
    \fmfv{label=$\vec n$}{A1}
    \fmfv{label=$\vec n$}{A3}
    \fmfv{label=$\vec m$}{A2}
    \fmfv{label=$\vec m$}{A4}
  \end{fmfchar*}}}\qquad
\parbox{30mm}{\fmfframe(5,5)(5,5){%
  \begin{fmfchar*}(20,20)
    \fmftop{A1,A2} \fmfbottom{A3,A4} 
    \fmf{photon}{A1,a}         
    \fmf{photon}{A2,a}
    \fmf{photon,label=$\nofrac{\vec k_+}{\vec k_-}$}{a,b} 
    \fmf{photon}{A3,b} 
    \fmf{photon}{A4,b}
    \fmfdot{a}
    \fmfdot{b}  
    \fmfv{label=$\vec n$}{A1}
    \fmfv{label=$\vec n$}{A3}
    \fmfv{label=$\vec m$}{A2}
    \fmfv{label=$\vec m$}{A4}
  \end{fmfchar*}}}\qquad
\parbox{30mm}{\fmfframe(5,5)(5,5){%
  \begin{fmfchar*}(20,20)
    \fmftop{A1,A2} \fmfbottom{A3,A4} 
    \fmf{photon}{A1,a} 
    \fmf{photon}{A3,b} 
    \fmf{photon,label=$\nofrac{\vec k_+}{\vec k_-}$}{a,b} 
    \fmf{phantom}{A4,b}
    \fmf{phantom}{A2,a}
    \fmffreeze
    \fmf{photon}{A4,a}
    \fmf{photon}{A2,b}
    \fmfdot{a}
    \fmfdot{b}  
    \fmfv{label=$\vec n$}{A1}
    \fmfv{label=$\vec n$}{A3}
    \fmfv{label=$\vec m$}{A2}
    \fmfv{label=$\vec m$}{A4}
  \end{fmfchar*}}}\quad
\parbox{25mm}{%
  \fmfframe(5,5)(5,5){%
    \begin{fmfchar*}(15,15)
      \fmftop{A1,A2} \fmfbottom{A3,A4} 
      \fmf{photon}{A1,a} 
      \fmf{photon}{A3,a} 
      \fmf{photon}{A4,a}
      \fmf{photon}{A2,a}
      \fmfdot{a}
      \fmfv{label=$\vec n$,la.di=5}{A1,A3}
      \fmfv{label=$\vec m$,la.di=5}{A2,A4}
    \end{fmfchar*}}}
\end{equation}
Compared to the simpler case discussed above, the Higgs
couplings~\eqref{eq:higgs-coupling} to gauge bosons with different
KK-quantum numbers are nonvanishing, so in the $t$- and $u$-channel
also the exchange of two physical scalars contributes:
\begin{equation}
\parbox{30mm}{\fmfframe(5,5)(5,5){%
  \begin{fmfchar*}(20,20)
    \fmftop{A1,A2} \fmfbottom{A3,A4} 
    \fmf{photon}{A1,a}         
    \fmf{photon}{A2,a}
    \fmf{dashes,label=$\nofrac{\vec k_+}{\vec k_-}$}{a,b} 
    \fmf{photon}{A3,b} 
    \fmf{photon}{A4,b}
    \fmfdot{a}
    \fmfdot{b}  
    \fmfv{label=$\vec n$}{A1}
    \fmfv{label=$\vec n$}{A3}
    \fmfv{label=$\vec m$}{A2}
    \fmfv{label=$\vec m$}{A4}
  \end{fmfchar*}}}\qquad
\parbox{30mm}{\fmfframe(5,5)(5,5){%
  \begin{fmfchar*}(20,20)
    \fmftop{A1,A2} \fmfbottom{A3,A4} 
    \fmf{photon}{A1,a} 
    \fmf{photon}{A3,b} 
    \fmf{dashes,label=$\nofrac{\vec k_+}{\vec k_-}$}{a,b} 
    \fmf{phantom}{A4,b}
    \fmf{phantom}{A2,a}
    \fmffreeze
    \fmf{photon}{A4,a}
    \fmf{photon}{A2,b}
    \fmfdot{a}
    \fmfdot{b}  
    \fmfv{label=$\vec n$}{A1}
    \fmfv{label=$\vec n$}{A3}
    \fmfv{label=$\vec m$}{A2}
    \fmfv{label=$\vec m$}{A4}
  \end{fmfchar*}}}
\end{equation}
\end{subequations}
The explicit form of the Higgs couplings are given by
\begin{equation}
\label{eq:higgs-example}
\begin{aligned}
  g_{HAA,\vec k_+,\vec n,\vec m}^{abc}&= -g f^{abc}\frac{(n_1-n_2)}{R}\\
  g_{HAA,\vec k_-,\vec n,\vec m}^{abc}&= -g f^{abc}\frac{(n_1+n_2)}{R}
\end{aligned}
\end{equation} 
In contrast to the scattering processes involving the same initial and final
KK-states, there is no contribution from the zero mode in the $t$- and 
$u$-channel. Since the cancellation of the unitarity violating terms
in~\eqref{eq:4n-sr} relies on the different couplings of the zero mode
and the KK-modes, this mechanism cannot be sufficient any more and the
physical scalars are expected to play an important role.

To take the contributions from the additional physical scalars into
account, the SR~\eqref{eq:4n-sr} has to be
modified. From~\eqref{eq:gold-lie} we get (exchanging
$a\leftrightarrow d$ and using that all initial and final state masses
are the same):
\begin{multline}
  \sum_{\vec k=\vec k_\pm}
     -\left(m_{\vec k}^2 g^{bde}_{\vec n,\vec m,\vec k}
     g^{ace}_{\vec n,\vec m,\vec k}+g_{HWW,\vec n,\vec m,\vec k}^{ebd}
     g_{HWW,\vec n,\vec m,\vec k}^{eac}\right)\\
  + \left(m_{\vec k}^2g^{bce}_{\vec n,\vec m,\vec k}g^{ade}_{\vec n,\vec m,\vec k}
  + g_{HWW,\vec n,\vec m,\vec k}^{ebc}g_{HWW,\vec n,\vec m,\vec k}^{ead}\right)
  = -4m_{\vec n}^2g^{abe}_{\vec n,\vec n,0}g^{cde}_{\vec m,\vec m,0}
\end{multline}
Inserting the coupling constants of the Higgs
bosons~\eqref{eq:higgs-example} and of the gauge
bosons~\eqref{eq:6d-3w} this turns into
\begin{multline}
  -2\frac{g^2}{R^2}\left[(n_1+n_2)^2+(n_1-n_2)^2\right]
      \left(f^{bde} f^{ace}-f^{bce}f^{ade}\right)\\
    = -4\frac{g^2}{R^2}(n_1^2+n_2^2)f^{abe}f^{cde}
\end{multline}
This relation is guaranteed by the Jacobi Identity.  Therefore the
unitarity cancellations in scattering among different KK-states at the
same mass-level take place because of an interplay of the KK-gauge
bosons and the `geometric Higgs bosons' with the peculiar form of the
coupling~\eqref{eq:higgs-coupling}.

\subsection{Sum rules for inelastic scattering}
We now turn to inelastic scattering at different KK-levels $\vec n$ to
$\vec m$.  Unlike the elastic case, no general SR for inelastic
scattering has been derived in~\cite{Csaki:2003dt} for a 5-dimensional
gauge theory.

The contributing diagrams are the same as in~\eqref{eq:4n-diag} with
the KK-momenta now given by
\begin{equation}
\label{eq:kk-vectors}
  \vec n=\binom{n_1}{n_2}\,,\,
  \vec m=\binom{m_1}{m_2}\,,\,
  \vec k_+=\binom{n_1+m_1}{n_2+m_2}\,,\,
  \vec k_-=\pm\binom{(n_1-m_1)}{(n_2-m_2)}
\end{equation}
where the sign in $\vec k_-$ has to be chosen so that $\pm
(n_1-m_1)\geq 0$.

From~\eqref{eq:gold-lie} we find the SR
\begin{multline}
\label{eq:inelastic-sr}
  \sum_{\vec k=\vec k_\pm}
    - \left(\frac{(m_{\vec k}^2+m_{\vec n}^2-m_{\vec m}^2)^2}
                 {m_{\vec k}^2} g^{bde}_{\vec n,\vec m,\vec k}
            g^{ace}_{\vec n,\vec m,\vec k}
          + g_{HWW,\vec n,\vec m,\vec k}^{ebd}
            g_{HWW,\vec n,\vec m,\vec k}^{eac}\right)\\
    + \left(\frac{(m_{\vec k}^2+m_{\vec n}^2-m_{\vec m}^2)^2 }
                 {m_{\vec k}^2}g^{bce}_{\vec n,\vec m,\vec k}
            g^{ade}_{\vec n,\vec m,\vec k}
          + g_{HWW,\vec n,\vec m,\vec k}^{ebc}
            g_{HWW,\vec n,\vec m,\vec k}^{ead}\right)\\
   = -4m_{\vec n}^2g^{abe}_{\vec n,\vec n,0}g^{cde}_{\vec m,\vec m,0}
\end{multline}
The combinations of the masses appearing in the SR can be simplified 
using the relation
\begin{equation}
\label{eq:mass-rel}
   m_{\vec k_\pm}^2+m_{\vec n}^2-m_{\vec m}^2=\frac{2}{R^2}\vec n
     \cdot(\vec n\pm\vec m)
\end{equation}
The Higgs coupling~\eqref{eq:higgs-coupling} takes the form
\begin{equation}
\label{eq:haa-general}
  g_{HAA,\vec k_\pm\vec n,\vec m}^{abc}
    = \sqrt 2 gf^{abc} \left[\frac{n_2m_1-n_1 m_2}{R^2 m_{\vec k_\pm}}\right]
\end{equation}
Let us first discuss the simpler case $n_2=m_2=0$.  The Higgs
couplings vanish and the situation is similar to scattering in
5~dimensions. The SR turns into
\begin{multline}
  -2\frac{g^2}{R^2}
    \left[\frac{(n^2+nm)^2}{(m+n)^2}+\frac{(n^2-nm)^2}{(m-n)^2}\right]
    \left(f^{bde} f^{ace}-f^{bce}f^{ade}\right)\\
  = - 4 \frac{g^2}{R^2}n^2f^{abe}f^{cde}
\end{multline}
Again this is guaranteed by the Jacobi Identity.

The required cancellation for the general
KK-momenta~\eqref{eq:kk-vectors} is less obvious and requires to take
the physical scalars into account.  Using the
expression~\eqref{eq:haa-general} for the Higgs coupling, one can
verify the relation
\begin{multline}
    \frac{(m_{\vec k_\pm}^2+m_{\vec n}^2-m_{\vec m}^2)^2 }
         {m_{\vec k_\pm}^2} g^{bde}_{\vec n,\vec m,\vec k}
    g^{ace}_{\vec n,\vec m,\vec k}
  + g_{HWW,\vec n,\vec m,\vec k_\pm}^{ebd}
    g_{HWW,\vec n,\vec m,\vec k_\pm}^{eac}\\
  = f^{bde}f^{ace}\frac{2g^2}{R^2}\vec n^2
\end{multline}
Inserting this result into the SR~\eqref{eq:inelastic-sr}, the
cancellation goes through as above.

Our discussion of gauge boson scattering shows that the physical
scalar components play an important role in the cancellation, iff the
initial and final state KK-momenta are not `parallel', a situation
that is a new phenomenon for gauge theories with more than one extra
dimension.

\section{Consistency of incomplete multiplets on the boundaries}
\label{sec:brane}

An important ingredient of orbifold GUT models is the explicit
symmetry breaking by matter on the orbifold fixed points, transforming
under the unbroken subgroup alone. A consistency check of such a setup
is provided by unitarity in the production of gauge bosons
corresponding to broken generators. This has been investigated
in~\cite{Hall:2001tn} for boundary Higgs bosons in 5-dimensional
$\mathrm{SU}(5)$ GUTs. In this example, the required cancellations
occur between the KK-zero-mode and the first KK-level and rely on an
apparent `conspiracy' among the coupling constants.  Here we provide a
general analysis for boundary fermions (other boundary fields can be
treated similarly) that shows that such cancellations are ensured by
the completeness relations of the KK-wavefunctions and the vanishing
wavefunctions of the broken gauge bosons on the fixed points.

\subsection{Matter on the boundaries}

The symmetry at the orbifold fixed points can be viewed as a
restricted gauge symmetry~\cite{Hall:2001pg}, since the gauge
parameters associated to the broken generators vanish on the boundary.
The BRST-transformations on the boundary at $y_f$ are therefore
\begin{subequations}
\label{eq:brane-brs}
\begin{align}
  \delta_{\text{BRST}} A^a_{\mu}(x,y_f)&=\partial_\mu c^a(x,y_f)
   +f^{abc}A_{\mu}^b(x,y_f) c^c(x,y_f)\\
  \delta_{\text{BRST}} \Phi^{\hat a}_{i}(x,y_f)&=\partial_i c^{\hat a}(x,y_f)
   + f^{\hat a\hat bc}\Phi_{i}^{\hat b}(x) c^c(x,y_f)
\end{align}
\end{subequations}
On the boundary, the gauge bosons only transform under the unbroken
subgroup, while the higher dimensional components of the gauge bosons
transform homogeneously under the unbroken subgroup and receive a
`shift' $\partial_i c^{\hat a}$ under the broken transformations. This
symmetry prohibits brane-mass terms for the $\Phi_i$ in 5 dimensions
and constrains the possible forms in
6~dimensions~\cite{vonGersdorff:2002rg}.  For symmetry breaking with
mixed BCs, the structure of the transformations on the boundary is
more complicated because the broken gauge (and ghost) fields are also
nonvanishing at the boundary and the separation of broken and unbroken
gauge transformations is lost.

On the boundaries we can add (possibly chiral) matter transforming
under the unbroken subgroup with generators $\tau_{L/R}$ satisfying
the Lie algebra~\eqref{eq:fermion-lie}.  The lagrangian for fermions
on the boundary is
\begin{equation}
\label{eq:ferm-brane}
  \begin{aligned}
    \mathscr{L}_f
      =&   \ii\bar\psi_i\fmslash\partial\psi
         + \bar\psi_i\fmslash A_a(y_f)(\tau^{a}_{Lij}(\tfrac{1-\gamma^5}{2})
            + \tau_{Rij}^a(\tfrac{1+\gamma^5}{2})\psi_j \\
      =&   \ii\bar\psi_i\fmslash\partial\psi
         + \bar\psi_i\fmslash A_\alpha(\mathcal{T}^{\alpha}_{Lij}(\tfrac{1-\gamma^5}{2})
            + \mathcal{T}_{Rij}^\alpha(\tfrac{1+\gamma^5}{2})\psi_j
  \end{aligned}
\end{equation}
with 
\begin{equation}\label{eq:t_aij}
  \mathcal{T}^{\alpha}_{L/R}=\tau^{a}_{L/R}f^\alpha(y_f)
\end{equation}
It is not possible to add Yukawa-Interactions of brane fermions to the 
scalar components of the bulk gauge bosons, since this violates the 
shift-symmetry $\propto \partial_i c^{\hat a}$ from~\eqref{eq:brane-brs}. 
To use a component of the higher dimensional gauge 
fields as Higgs boson, one either has to put the fermions in the 
bulk or introduce a non-local coupling~\cite{Csaki:2002ur} that 
can be generated  from mixing with bulk fermions.

The lagrangian of the KK-modes is invariant under the BRST
transformations~\eqref{eq:brs} and
\begin{equation}
  \delta_{\text{BRST}} \psi_{L/R\,i}
    = \ii c^\alpha(x) \mathcal{T}^{\alpha}_{L/R\,ij}\psi_{L/R\,j}
\end{equation}
iff the relation 
\begin{equation}
\label{eq:ferm-sr}
  [\mathcal{T}^\alpha_{L/R},\mathcal{T}^\beta_{L/R}]
    = g^{\alpha\beta\gamma}\mathcal{T}^\gamma_{L/R}
\end{equation}
is satisfied, which also ensures the nilpotency of the BRST
transformation.  We will discuss this equation below, where the same
relation appears as condition for unitarity.  It should be noted that
on the right hand side also the coupling constants $g^{\hat \alpha
\hat \beta \gamma}$ with broken indices can appear, in contrast to the
BRST transformations restricted to the brane~\eqref{eq:brane-brs}. In
the KK-picture the broken indices appear because the KK-modes are not
localized in the extra dimension and the reduced symmetry is not
apparent.

\subsection{Unitarity in gauge boson production}

We turn to the SRs for scattering of brane fermions into gauge bosons.
For the scattering into unbroken gauge bosons, the KK-excitations can
couple to the brane fermions and to an intermediate unbroken gauge
boson, so that the contributing diagrams are
\begin{equation}
  \parbox{20mm}{%
    \begin{fmfchar*}(20,20)
      \fmftop{A1,A2} \fmfbottom{A3,A4} 
      \fmf{fermion}{A1,a} 
      \fmf{photon}{A2,b}
      \fmf{photon,label=$\gamma$}{a,b} 
      \fmf{fermion}{a,A3} \fmf{photon}{A4,b}
      \fmfdot{a}
      \fmfdot{b}
      \fmfv{label=$\beta$,label.angle=0}{A4}
      \fmfv{label=$\alpha$,label.angle=0}{A2}  
    \end{fmfchar*}}\qquad\qquad
  \parbox{20mm}{%
    \begin{fmfchar*}(20,20)
      \fmftop{A1,A2} \fmfbottom{A3,A4} 
      \fmf{fermion}{A1,a} 
      \fmf{photon}{A2,a}
      \fmf{fermion}{a,b} 
      \fmf{fermion}{b,A3} 
      \fmf{photon}{A4,b}
      \fmfv{label=$\beta$,label.angle=0}{A4}
      \fmfv{label=$\alpha$,label.angle=0}{A2}  
      \fmfdot{a}
      \fmfdot{b}  
    \end{fmfchar*}}\qquad\qquad
  \parbox{20mm}{%
    \begin{fmfchar*}(20,20)
      \fmftop{A1,A2} \fmfbottom{A3,A4} 
      \fmf{fermion}{A1,a} 
      \fmf{fermion}{b,A3} 
      \fmf{fermion}{a,b} 
      \fmf{phantom}{A4,b}
      \fmf{phantom}{A2,a}
      \fmffreeze
      \fmf{photon}{a,A4}
      \fmf{photon,rubout}{A2,b}
      \fmfdot{a}
      \fmfdot{b}
      \fmfv{label=$\beta$,label.angle=0}{A4}
      \fmfv{label=$\alpha$,label.angle=0}{A2}  
    \end{fmfchar*}}
\end{equation}
The SR obtained from the cancellation of the leading divergences (or
equivalently the WI with a single contraction) is just the Lie
algebra~\eqref{eq:ferm-sr}.  Inserting the
definitions~\eqref{eq:g_abc} and~\eqref{eq:t_aij}, we find that this
relation is satisfied because of the completeness of the
KK-wavefunctions and the Lie algebra~\eqref{eq:fermion-lie}.
Explicitly, 
the RHS becomes
\begin{equation}
\label{eq:fermion-simplify}
\begin{aligned}
   g^{\alpha\beta\gamma}\mathcal{T}^\gamma_{L/R}
    &= f^{abc}\tau^c_{L/R} \sum_{\gamma} f^\gamma(y_f)
         \int\!\dd^Ny\,f^\alpha(y) f^\beta(y) f^\gamma(y)\\
    &= f^{abc}\tau^c_{L/R} f^\alpha(y_f) f^\beta(y_f)
\end{aligned}
\end{equation}
Thus the same KK-wavefunctions appear as on the left hand side
and~\eqref{eq:ferm-sr} is reduced to the simple Lie algebra of the
$\tau^a_{L/R}$.

Only the $s$-channel diagrams contribute to the production of broken
gauge bosons, since the boundary fermions completely decouple from the
broken gauge bosons:
\begin{equation}
  \parbox{30mm}{%
    \fmfframe(5,5)(5,5){%
      \begin{fmfchar*}(20,20)
        \fmftop{A1,A2} \fmfbottom{A3,A4} 
        \fmf{fermion}{A1,a} 
        \fmf{photon}{A2,b}
        \fmf{photon,label=$\gamma$}{a,b} 
        \fmf{fermion}{a,A3}
        \fmf{photon}{A4,b}
        \fmfdot{a}
        \fmfdot{b}
        \fmflabel{$\hat \alpha$}{A2}  
        \fmflabel{$\hat \beta$}{A4}  
      \end{fmfchar*}}}
\end{equation}
Therefore the SR~\eqref{eq:ferm-sr} reduces to 
\begin{equation}
\label{eq:ferm-sr-broken}
  g^{\hat\alpha\hat\beta\gamma}\mathcal{T}^\gamma_{L/R}=0\,.
\end{equation}
Exploiting the completeness relation of the KK-wavefunctions once
more, we see that this relation is indeed satisfied because the broken
wavefunctions vanish on the boundary:
\begin{equation}
\begin{aligned}
  g^{\hat\alpha\hat\beta\gamma}\mathcal{T}^\gamma_{L/R}
    &= f^{\hat a\hat bc}\tau^c_{L/R} \sum_{\gamma} f^{\gamma}(y_f)
          \int\!\dd^Ny\, f^{\hat \alpha}(y) f^{\hat \beta}(y) f^{\gamma}(y)\\
    &= f^{\hat a\hat bc}\tau^c_{L/R} f^{\hat \alpha}(y_f) f^{\hat\beta}(y_f) = 0
\end{aligned}
\end{equation}
For massive chiral fermions, subleading divergences appear whose
cancellation requires the SR~\eqref{eq:2f2w-wi2}. Masses of the chiral
brane-fermions are not included in our present setting and have to be
generated by a further breaking of the unbroken subgroup, e.\,g.~by a
Higgs boson localized on the brane. The investigation of such
constructions is beyond the scope of the present work.

\section{Summary and outlook}

We have performed the KK-decomposition of a general gauge theory on an
$C/Z_2^n$ orbifold and determined consistent boundary conditions that
allow BRST quantization and to derive WIs. This yields a new
demonstration of the consistency of orbifold symmetry breaking and of
the Dirichlet boundary conditions currently employed for models of
EWSB without Higgs bosons~\cite{Csaki:2003dt,Csaki:2003zu}.  On the
other hand, mixed boundary conditions turn out to be inconsistent
without the introduction of a Higgs multiplet on the boundary.  In the
presence of such a brane Higgs multiplet, the GBs are a mixture of the
unphysical components of the boundary Higgs with the higher
dimensional components of the gauge bosons~\cite{Mueck:2001}.
Dirichlet BCs arise from the limit of an infinite vacuum expectation
value of a boundary Higgs.  As an extension of our work, the
consistency of such a setup could be further clarified by quantizing
such theories including the boundary Higgs, taking the limit $v\to
\infty$ at the end.
 
Recently, a Higgsless mechanism has also been proposed for the
generation of fermion masses by BCs~\cite{Csaki:fermions}.  The
application of the approach presented here to this construction is
given elsewhere~\cite{CS:fermions}.

In section~\ref{sec:higgs-sr} we have shown that on tree level the
structure of the lagrangian of a SBGT is fixed by imposing a finite
set of WIs for 4-point functions without external goldstone bosons.  The
conditions derived from the WIs include the unitarity-SRs derived from
tree level unitarity
in~\cite{LlewellynSmith:1973,Cornwall:1973,Gunion:1991}.  The same SRs
are also valid in compactified higher dimensional gauge theories
broken by orbifold BCs, provided the sum over the KK-tower converges.
The introduction of a cutoff or regularization of the KK-sums can
upset gauge invariance and requires further considerations.

In section~\ref{sec:kk-sr} the SRs have been applied to a
6-dimensional gauge theory. We have shown that the physical scalar
components of the gauge bosons play an important role in ensuring the
unitarity cancellations in gauge boson scattering where the final
state KK-momenta are not parallel to that of the initial state gauge
bosons. The clarification of this mechanism should prove useful in
extending the discussion of partial wave unitarity in
KK-theories~\cite{SekharChivukula:2001hz,DeCurtis:2002} to
6-dimensional models.  In section~\ref{sec:brane} we have demonstrated
the consistency of placing reduced multiplets at the orbifold fixed
points.

After this work was completed, a discussion of unitarity in the
Higgs\-less models of~\cite{Csaki:2003zu} appeared~\cite{Rizzo}, where
it is pointed out that partial wave unitarity might be violated at a
scale of $\sqrt s\sim 2\, \mathrm{TeV}$ despite the fulfillment of the
unitarity sum rules if the mass of the first KK-excitation is too
large. This poses a problem for the warped version of the model while
in a flat space model no such problems have been found.  While the
present work was concerned with gauge theories on separable background
metrics, our approach will also be useful for studying unitarity of
gauge boson scattering in a warped background and for answering open
questions.

\subsection*{Acknowledgments}
We thank A.~M\"uck for useful discussions. This work has been
supported by the Bundesministerium f\"ur Bildung und Forschung of
Germany, grant 05HT1RDA/6. C.\,S. is supported by the Deutsche
Forschungsgemeinschaft through the Gra\-du\-ier\-ten\-kolleg
`Eichtheorien' at Mainz University.

\appendix
\section{Kaluza-Klein lagrangian and couplings}
\label{app:kk-lag}

The $(4+N)$-dimensional Yang Mills lagrangian is
\begin{equation}
   \mathscr{L}_{4+N}=-\frac{1}{4} F^a_{AB}(x,y)F^{a,AB}(x,y)
\end{equation}
with the field strength
\begin{equation}
  F^a_{AB}(x,y)=\partial_A A^a_B(x,y)
     - \partial_B A^a_A(x,y) + f^{abc}A^b_A(x,y)A^c_B(x,y)
\end{equation}
Here we include the higher dimensional gauge coupling $g_D$ in the
structure constants. The BRST transformations are
\begin{equation}
\label{eq:d-brs}
  \begin{aligned}
    \delta_{\text{BRST}} A^a_A(x,y)&=\partial_A c^a(x,y)+f^{abc}A_A^b(x,y) c^c(x,y)\\
    \delta_{\text{BRST}} c^a(x,y)&=-\frac{1}{2}f^{abc}c^b(x,y)c^c(x,y)\\
    \delta_{\text{BRST}} \bar c^a(x,y)&=B^a(x,y)\\
    \delta_{\text{BRST}} B^a(x,y)&=0
  \end{aligned}
\end{equation}
where the equation of motion of the auxiliary field $B^a$ is
\begin{equation}
  B^a = -\frac{1}{\xi}G^a
      = -\frac{1}{\xi}(\partial_\mu A^{a,\mu}(x,y)-\xi \partial_i\Phi^{a,i}(x,y))
\end{equation} 
The complete lagrangian of the KK-modes is
\begin{equation}
\label{eq:kk-lag}
  \begin{aligned}
    \mathscr{L}_{KK}
     &= - \frac{1}{4}(\partial_\mu A_{\nu}^\alpha-\partial_\nu A_{\mu}^\alpha)
                     (\partial^\mu A^{\alpha,\nu}-\partial^\nu A^{\alpha,\mu})
        + \frac{1}{2} m_{\alpha}^2A^\alpha_{\mu}A^{\alpha,\mu}\\
     &  - m_{\alpha_i}\partial_\mu \Phi^{\alpha,i} A^{\alpha,\mu}
        - g^{\alpha\beta\gamma}\partial_\mu A^\alpha_\nu A^{\beta,\mu} A^{\gamma,\nu} 
        - \frac{1}{4}g^{\alpha\beta\gamma\delta}
            A^\alpha_{\mu}A^{\beta}_\nu A^{\gamma,\mu}A^{\delta,\nu} \\
     &  + \frac{1}{2}\partial_\mu \Phi^{\alpha}_i \partial^\mu \Phi^{\alpha,i}
        - \frac{1}{2}\left[m_{\alpha}^2 \Phi^\alpha_{i} \Phi^{\alpha,i}
        - (m_{\alpha_i}\Phi^{\alpha,i})^2\right]\\
     &  - \frac{1}{2}T^{\alpha}_{\beta\gamma}\, A^{\alpha,\mu}
            \Phi^{\beta}_i\overleftrightarrow{\partial_\mu} \Phi^{\gamma,i}
        + \frac{1}{2}g_{\Phi AA}^{i\alpha\beta\gamma}
            \Phi^{\alpha,i} A_{\mu}^{\beta}A^{\mu,\gamma}
        + \frac{1}{4}g_{A^2\Phi^2}^{\alpha\beta\gamma\delta}
            A^\alpha_{\mu}A^{\beta}_{\nu} \Phi^{\gamma,i}\Phi^{\delta}_i\\
     &  - \frac{1}{2} T^{\alpha}_{\beta\gamma}
            (m_{\alpha_j}\Phi^\alpha_{i}-m_{\alpha_i}\Phi^\alpha_{j})
               \Phi^{\beta,i}\Phi^{\gamma,j}
        - \frac{1}{4}g_{\Phi^4}^{\alpha\beta\gamma\delta}
            \Phi^\alpha_{i}\Phi^\beta_{j}\Phi^{\gamma,i}\Phi^{\delta,j}
  \end{aligned}
\end{equation}
with the coupling constants defined in~\eqref{eq:kk-couplings} and 
\begin{subequations}
\label{eq:kk-couplings-app}
\begin{align}
  g^{\alpha\beta\gamma\delta}
    &= f^{abe}f^{cde}\int\!\dd^N y\, f^\alpha (y) f^\beta(y)f^\gamma(y)f^\delta(y)\\
\label{eq:ti_abc}
  g_{\Phi AA}^{i\alpha\beta\gamma}
    &= m_{\beta_i}T^\gamma_{\beta\alpha} + m_{\gamma_i}T^\beta_{\gamma\alpha}\\
  g_{A^2\Phi^2}^{\alpha\beta\gamma\delta}
    &= 2f^{abe}f^{cde}\int\!\dd^N y \,f^\alpha (y)f^\beta (y)g^\gamma (y)g^\delta(y)\\
  g_{\Phi^4}^{\alpha\beta\gamma\delta}
    &= f^{abe}f^{cde}\int\!\dd^N y \,g^\alpha (y) g^\beta(y)g^\gamma (y)g^\delta(y)
\end{align}
\end{subequations}

\section{Parameterization of the general lagrangian}
\label{app:general-fr}
We give our parameterization of the general lagrangian with the
particle spectrum of a SBGT used in the calculation of the WIs.
Apart from terms $\propto \epsilon^{\mu\nu\rho\sigma}W_\mu W_\nu
W_\rho W_\sigma$, the most general renormalizable interaction
lagrangian for these fields is
\begin{subequations}
\label{subeq:ssb-lag}
\begin{equation}
\label{eq:ssb-lag}
  \begin{aligned}
    \mathscr{L}_{\text{int}}
      =& - f^{abc}W_{b,\mu} W_{c,\nu}\partial^\mu W_a^\nu
         - \frac{1}{4}g_{W^4}^{abcd} W_{a,\mu} W_{b,\nu} W_c^\mu W_d^\nu
         - \frac{1}{2}t_{ab}^c(\phi_a\overleftrightarrow{\partial_\mu}\phi_b) W_c^\mu\\
       & + \frac{g_{\phi WW}^{a bc}}{2}\phi_a W_b^\mu W_{c,\mu}
         - \frac{1}{2}T^a_{ij}(H_i\overleftrightarrow{\partial_\mu}H_j)W_a^\mu
         + g_{H\phi W}^{ia b}(\phi_a\overleftrightarrow{\partial_\mu}  H_i) W_b^\mu\\
       & + \frac{1}{2}g_{HWW}^{iab}H_iW_{a,\mu}W_b^\mu
         + \frac{1}{4}g_{\phi^2 W^2}^{ab cd}\phi_a\phi_b
           W_{c,\mu}W_d^\mu+\frac{1}{4}g_{H^2 W^2}^{ab ij}H_iH_jW_a^\mu W_{b,\mu}\\
       & + \frac{1}{2}g_{H\phi W^2}^{a bc i} H_i\phi_a W_{b,\mu}W_c^\mu
         + \frac{1}{2}g_{\phi^2 H}^{ab i}\phi_a\phi_b H_i
         + \frac{1}{2}g_{\phi H^2}^{a ij} \phi_a H_iH_j\\
       & + \frac{1}{3!}g_{\phi^3}^{abc}\phi_a\phi_b\phi_c
         + \frac{1}{3!}g_{H^3}^{ijk}H_iH_jH_k 
         + \text{quartic scalar interactions}
  \end{aligned}
\end{equation}
The lagrangian of the fermions is parametrized by 
\begin{multline}
\label{eq:ferm-ssb-lag}
  \mathscr{L}_f
    =   \ii\bar\psi_i\fmslash\partial\psi
      + \bar\psi_i\fmslash W_a(\tau^{a}_{Lij}(\tfrac{1-\gamma^5}{2})
      + \tau_{Rij}^a(\tfrac{1+\gamma^5}{2})\psi_j \\
      + \bar\psi_i \phi_a (g_{\phi ij}^a(\tfrac{1-\gamma^5}{2})
      + g_{\phi ij}^{a\dagger}(\tfrac{1+\gamma^5}{2}))\psi_j 
      + \bar\psi_i H_k (g_{H ij}^k(\tfrac{1-\gamma^5}{2})
      + g_{Hij}^{k\dagger} (\tfrac{1+\gamma^5}{2}))\psi_j
\end{multline}
\end{subequations}
This can be compared with the lagrangian of a SBGT
\begin{equation}
  \begin{aligned}
    \mathscr{L}
      &= - \frac{1}{4}F^{\mu\nu}_{a}F_{a\mu\nu}
         + \frac{1}{2}D_\mu\phi_AD^\mu\phi_A-V(\phi)\\
      &  + \ii\bar\psi_i\fmslash D\psi_i 
         + \bar\psi_i \phi_A \bigl(X_{ij}^A(\tfrac{1-\gamma^5}{2})
         + {X_{ij}^A}^\dagger(\tfrac{1+\gamma^5}{2})\bigr)\psi_j
  \end{aligned}
\end{equation}
with 
\begin{align}
  F_{a\mu\nu} &= \partial_\mu A_{a,\nu} - \partial_\nu A_{a,\mu} 
       + f^{abc}A_{b,\nu}A_{c,\nu}\\
\label{eq:phi-der}
  D_\mu\phi_A &= \partial_\mu\phi_A + T_{AB}^a W_{a,\mu}\phi_B\\
\label{eq:scalar-potential}
  V(\phi) &=   \frac{g_2^{AB}}{2}\phi^A\phi^B
             + \frac{g_3^{ABC}}{3!}\phi^A\phi^B\phi^C
             + \frac{g_4^{ABCD}}{4!}\phi^A\phi^B\phi^C\phi^D\\
  D_\mu\psi_i &=   \partial_\mu\psi_i
                 - \ii W_{a,\mu}(\tau^a_{Lij}(\tfrac{1-\gamma^5}{2})
                 + \tau_{Rij}^a(\tfrac{1+\gamma^5}{2}))\psi_j
\end{align}
Inserting the parameterization~\eqref{eq:t-def} for the generators in
the representation of the scalars, we find that the definitions of
$T_{ij}^a$, $t_{ab}^c$ and $g_{H\phi W}$ agree with those in the
lagrangian~\eqref{eq:ssb-lag}.  The 2 scalar-2 gauge boson coupling is
given by the anticommutator of representation matrices:
\begin{equation}
  g_{\phi^2 W^2}^{ABcd}=-\{T^c,T^d\}_{AB}
\end{equation}
The cubic scalar gauge boson couplings originate from the contraction
of the anticommutator $\{T^a,T^b\}$ with a vacuum expectation value
$\phi_0$:
\begin{equation}
\label{eq:phiww-def}
  g_{\phi WW}^{Abc}=g_{\phi^2 W^2}^{ABbc}\phi_{0B}
\end{equation}
The triple scalar couplings can be expressed through the terms in the
scalar potential by
\begin{equation}
\label{eq:cubic-higgs}
  g_{\Phi^3}^{ABC}=- g_{3}^{ABC}-g_4^{iABC}v_i
\end{equation}

\section{Form of the Goldstone boson couplings}
\label{app:gb-couplings}

The relations~\eqref{subeq:unitarity-relations} follow from the condition
that the gauge transformation of the vacuum expectation value must be
in the Goldstone boson direction
\begin{equation}
\label{eq:diag_mass}
  T^a_{bc}\phi_{0c}
   = \begin{pmatrix}
       m_{W_a}\\
       0
     \end{pmatrix} \delta_{a,b}
\end{equation}
Using~\eqref{eq:phiww-def} this implies
\begin{equation}\label{eq:gphiww}
  g_{\phi WW}^{Abc}=(m_b T^c_{bA}+m_c T^b_{cA})
\end{equation}
Acting on the vacuum expectation value $\phi_0$ with a commutator of
two generators in the representation of the scalars results in:
\begin{equation}
\label{eq:consistent-lie}
   \left([\mathbf{T}^a,\mathbf{T}^b]\right)\phi_0
     = f^{abc} \begin{pmatrix}
                 m_{W_c} \\
                 0
               \end{pmatrix}
    \Rightarrow 
    \begin{cases}
     & t^a_{cb}m_{W_b}-t^b_{ca}m_{W_a} = f^{abc}m_{W_c}\\
     &m_a g^{iab}_{H\phi W}-m_b g^{iba}_{H\phi W}=0
    \end{cases}
\end{equation}
Together with~\eqref{eq:gphiww}, the first relation implies the
relation~\eqref{eq:2phi-w-coupling}~\cite{Schwinn:2003} while the second one
implies~\eqref{eq:hphiw-coupling} .  Similarly,
contracting the transformation law of the Yukawa
couplings~\eqref{eq:yukawa-tensor} with the vacuum expectation value
$\phi_{0A}$ and using the condition that the fermions get their masses
from the coupling to the scalars
\begin{equation}
\label{eq:yukawa-ssb-condition}
  X_{ij}^A\phi_{0A} = {X_{ij}^A}^\dagger\phi_{0A} =-\delta_{ij}m_i
\end{equation}
one can derive the fermion-Goldstone boson
coupling~\eqref{eq:f-phi-coupling}.

The remaining relations on the cubic GB
couplings~\eqref{eq:g_phi_2h},~\eqref{eq:g_2phi_h} and~\eqref{eq:g_3phi}
are consequences of the invariance of
the scalar potential
\begin{equation}
\label{eq:potential-gauge}
  \frac{\partial V(\phi)}{\partial \phi_A}T^a_{AB}\phi_B=0
\end{equation}
as can be seen by taking two derivatives with respect to $\phi$ and
setting $\phi=\phi_0$.

\section{Results from the WIs}
\label{app:wi-results}
In this appendix we collect the results of the WIs for the 3- and
4-point functions with up to 4 contractions. In the calculations the
general lagrangian~\eqref{subeq:ssb-lag} has been used.  No symmetry
relations among the coupling constants have been assumed, but only
results of previously evaluated WIs have been used to simplify the
calculations.  The SRs obtained in this way allow to express all
coupling constants in terms of the input parameters
\begin{equation}
\label{eq:input}
  f^{abc}\,,\,
  \tau_{L/R\,ij}^a\,,\,
  g_{Hij}^h\,,\,
  g_{HWW}^{iab}\,,\,
  T^a_{ij}\,,\,
  g_{H^3}^{ijk}\,,\,
  g_{H^4}^{ijkl}
\end{equation}
This is not a \emph{minimal} set of input parameters since they are
subject to constraints arising from the Lie algebra, the Jacobi
Identities and other symmetry relations.  The detailed calculations of
the WIs can be found in~\cite{Schwinn:2003}.

\subsection{Cubic Goldstone boson couplings}
\label{app:reconstruct_3point}
The couplings of one Goldstone boson to two physical particles are
determined by the WIs with one contraction.
\begin{subequations}
\label{subeq:triple-phi-wi1}
\begin{align}
  \parbox{15mm}{%
    \begin{fmfchar}(15,15)
      \fmfleft{A1,A2}
      \fmfright{A3}
      \fmf{double}{A1,a}
      \fmf{dashes}{A2,a}
      \fmf{photon}{A3,a}
      \fmfblob{10}{a}
    \end{fmfchar}}=0&\qquad \Rightarrow
    g_{H\phi W}^{iab}=-\frac{1}{2m_{W_a}}g_{HWW}^{iab}\\
\label{eq:gold-ferm-wi}
  \parbox{15mm}{%
    \begin{fmfchar}(15,15)
      \fmfleft{A1,A2}
      \fmfright{A3}
      \fmf{double}{A1,a}
      \fmf{fermion}{A2,a}
      \fmf{fermion}{a,A3}
      \fmfblob{10}{a}
    \end{fmfchar}}=0 &\qquad \Rightarrow 
    g_{\phi ij}^a=-\frac{\ii}{m_{W_a}}(m_{f_i}\tau_{L ij}^a-m_{f_j}\tau_{R ij}^a)\\
\label{eq:g_phi_2h} 
 \parbox{15mm}{%
    \begin{fmfchar}(15,15)
      \fmfleft{A1,A2}
      \fmfright{A3}
      \fmf{double}{A1,a}
      \fmf{dashes}{A2,a}
      \fmf{dashes}{a,A3}
      \fmfblob{10}{a}
    \end{fmfchar}}=0 &\qquad \Rightarrow
    g_{\phi H^2}^{aij}=\frac{1}{m_{W_a}}T^a_{ij}(m_i^2-m_j^2) \\
  \parbox{15mm}{%
    \begin{fmfchar}(15,15)
      \fmfleft{A1,A2}
      \fmfright{A3}
      \fmf{double}{A1,a}
      \fmf{photon}{A2,a}
      \fmf{photon}{a,A3}
      \fmfblob{10}{a}
    \end{fmfchar}}=0 &\qquad \Rightarrow
    g_{\phi WW}^{abc}=\frac{1}{m_{W_a}}f^{abc}(m_{W_b}^2-m_{W_c}^2)
\end{align}
\end{subequations}
(in all diagrams, the insertion of the operator $(\partial_\mu
W^\mu-m_W\phi)$ is represented by a double line).
To obtain the relations for the couplings of 2 Goldstone bosons to one 
physical particle, one has to consider the WIs~\eqref{eq:chanowitz}
with two contractions
\begin{subequations}
\label{subeq:triple-phi-wi2}
\begin{align}
\label{eq:g_2phi_h}
  \parbox{15mm}{%
    \begin{fmfchar}(15,15)
      \fmfleft{A1,A2}
      \fmfright{A3}
      \fmf{double}{A1,a}
      \fmf{dashes}{A2,a}
      \fmf{double}{A3,a}
      \fmfblob{10}{a}
    \end{fmfchar}}=0&\quad\Rightarrow 
    g^{abi}_{\phi^2H}=-\frac{m_{H_i}^2}{2m_{W_a}m_{W_b}}g_{HWW}^{iab}\\
  \parbox{15mm}{%
    \begin{fmfchar}(15,15)
      \fmfleft{A1,A2}
      \fmfright{A3}
      \fmf{double}{A1,a}
      \fmf{photon}{A2,a}
      \fmf{double}{A3,a}
      \fmfblob{10}{a}
    \end{fmfchar}}=0 &\qquad \Rightarrow
    m_{W_a} m_{W_c} t^b_{a c}=\frac{1}{2}f^{b ac}(m_{W_b}^2-m_{W_a}^2-m_{W_c}^2)\\
\label{eq:g_3phi}
  \parbox{15mm}{%
    \begin{fmfchar}(15,15)
      \fmfleft{A1,A2}
      \fmfright{A3}
      \fmf{double}{A1,a}
      \fmf{double}{A2,a}
      \fmf{double}{A3,a}
      \fmfblob{10}{a}
    \end{fmfchar}}=0 &\qquad \Rightarrow
    g_{\phi^3}^{abc}=0
\end{align}
\end{subequations}

\subsection{Gauge couplings of physical particles}
\label{app:physical}
The Lie algebra structure of the couplings of the physical particles,
i.\,e.~the Higgs bosons, gauge bosons and fermions, arises from WIs for
4-point functions with one contraction, together with the quartic
gauge couplings and the $2W2H$ coupling:
\begin{align}
\label{eq:2f2w-wi}
  \parbox{20mm}{%
    \begin{fmfchar*}(20,20)
      \fmfleft{f1,f2}
      \fmfright{A,H}
      \fmf{fermion,label=$f_j$}{f1,a}
      \fmf{fermion,label=$f_i$, label.side=left}{a,f2}
      \fmf{double}{A,a}
      \fmf{photon,label=$W_b$,label.side=left}{H,a}     
      \fmfv{decor.shape=circle,decor.filled=empty,decor.size=25}{a}   
    \end{fmfchar*}} &\Rightarrow
    \begin{cases}
      [\tau_{L}^a, \tau_{L}^b]_{ij} - \ii f^{abc}\tau_{Lij}^c&=0\\
      [\tau_{R}^a, \tau_{R}^b]_{ij} - \ii f^{abc}\tau_{Rij}^c&=0
    \end{cases} \\[5mm]
  \parbox{20mm}{%
    \begin{fmfchar*}(20,20)
      \fmfleft{A1,A2}
      \fmfright{A3,A4} 
      \fmf{double}{A2,a} 
      \fmf{dashes,label=$H_i$,label.side=right}{A3,a}
      \fmf{photon,label=$W_b$}{A1,a} 
      \fmf{dashes,label=$H_j$,label.side=left}{A4,a}
      \fmfv{decor.shape=circle,decor.filled=empty,decor.size=25}{a}
    \end{fmfchar*}}&\Rightarrow
    \begin{cases}
      [T^a,T^b]_{ij}
       - [\tfrac{g_{HWW}^{a}}{2m_{W_c}},\tfrac{g_{HWW}^{b}}{2m_{W_c}}]_{ij}
         &= f^{abc}T^c_{ij}\\
      \{T^a_{ik},T^b_{kj}\}
       - \{\tfrac{g_{HWW}^{a}}{2m_{W_c}},\tfrac{g_{HWW}^{b}}{2m_{W_c}}\}_{ij}
         &= -g_{H^2 W^2}^{abij}
    \end{cases}\label{eq:2h2w-wi}
\end{align}

\subsubsection{Yukawa couplings}
The symmetry conditions of the Yukawa
couplings~\eqref{eq:yukawa-tensor} can be obtained from the WI for 2
fermions, one gauge boson and one Higgs boson and the WI for two
fermions and 2 contractions:
\begin{align}
\label{eq:2fhw-wi}
  \parbox{20mm}{%
    \begin{fmfchar*}(20,20)
      \fmfleft{f1,f2}
      \fmfright{A,H}
      \fmf{fermion,label=$f_j$}{f1,a}
      \fmf{fermion,label=$f_i$,label.side=left}{a,f2}
      \fmf{double}{A,a}
      \fmf{dashes,label=$H_h$,label.side=left}{H,a}
      \fmfv{decor.shape=circle,decor.filled=empty,decor.size=25}{a}
    \end{fmfchar*}} &\Rightarrow
  \begin{aligned}
    0=-\ii\frac{g_{HWW}^{hba}}{2m_{W_a}} g_{\phi ij}^a
      - \ii (g_{H ij}^h T^b_{hk}) - g_{H il}^k\tau_{Llj}^b+\tau_{Ril}^bg_{H lj}^k
  \end{aligned}\\[5mm]
\label{eq:2f2w-wi2}
  \parbox{20mm}{%
    \begin{fmfchar*}(20,20)
      \fmfleft{f1,f2}
      \fmfright{A,H}
      \fmf{fermion,label=$f_j$}{f1,a}
      \fmf{fermion,label=$f_i$,label.side=left}{a,f2}
      \fmf{double}{A,a}
      \fmf{double}{H,a}     
      \fmfv{decor.shape=circle,decor.filled=empty,decor.size=25}{a} 
    \end{fmfchar*}} &\Rightarrow
  \ii g_{\phi il}^b\tau_{Llj}^a  - \ii \tau_{Ril}^ag_{\phi lj}^b
    = g_{\phi ij}^c t^a_{cb}-g_{H ij}^k \frac{g_{HWW}^{kab}}{2m_{W_b}}
\end{align}

\subsection{Goldstone-gauge boson couplings}
\label{app:gold-couplings}
The components of the graded Jacobi Identity~\eqref{eq:2s-jac} are
reproduced by the the WI for the $3W H$ amplitude with one contraction
and the $4W$ WI with 2 contractions:
\begin{align}
\label{eq:hphi-jac-exp}
  \parbox{20mm}{%
    \begin{fmfchar*}(20,20)
      \fmfleft{A1,A2}
      \fmfright{A3,A4}
      \fmf{double}{A1,a}
      \fmf{dashes,label=$H_i$,label.side=right}{A2,a}
      \fmf{photon,label=$W_b$}{A3,a}
      \fmf{photon,label=$W_c$}{A4,a}
      \fmfv{decor.shape=circle,decor.filled=empty,decor.size=25}{a}
    \end{fmfchar*}} &\Rightarrow
  \begin{aligned}
    m_{W_a}g_{H\phi W^2}^{abci}
      =& -g_{HWW}^{icd}f^{abd}-g_{HWW}^{ibd}f^{acd}\\
       & -\frac{g_{HWW}^{iad}}{2m_{W_d}}g^{dbc}_{\phi WW}+T^a_{ij}g_{HWW}^{jbc}
  \end{aligned}\\[5mm]
\label{eq:2phi-jac-exp}
  \parbox{20mm}{%
    \begin{fmfchar*}(20,20)
      \fmfleft{A1,A2}
      \fmfright{A3,A4}
      \fmf{double}{A1,a}
      \fmf{photon,label=$W_c$,label.side=right}{A2,a}
      \fmf{double}{A3,a}
      \fmf{photon,label=$W_d$}{A4,a}
      \fmfv{decor.shape=circle,decor.filled=empty,decor.size=25}{a}
    \end{fmfchar*}} & \Rightarrow 
  \begin{aligned}
     m_{W_a}m_{W_b}g_{\phi^2 W^2}^{abcd} - \frac{1}{2}g_{HWW}^{iab}g_{HWW}^{icd}
       = m_{W_b}g_{\phi WW}^{ecd}t^a_{be}\\
          + f^{ace}f^{dbe}(m_{W_d}^2-m_{W_e}^2) - f^{cbe}f^{dae}(m_{W_c}^2-m_{W_e}^2)
  \end{aligned}
\end{align}
The $ia$  component of the Lie algebra~\eqref{eq:scalar-lie}
follows from the $3WH$ WI with two contractions:
\begin{equation}
\label{eq:3wh-wi2}
  \parbox{20mm}{%
    \begin{fmfchar}(20,20)
      \fmfleft{A1,A2}
      \fmfright{A3,A4}
      \fmf{double}{A1,a}
      \fmf{double}{A2,a}
      \fmf{dashes}{A3,a}
      \fmf{photon}{A4,a}
      \fmfv{decor.shape=circle,decor.filled=empty,decor.size=25}{a}
    \end{fmfchar}}\Rightarrow
  \begin{aligned}
    - \frac{1}{2m_{W_e}^2} g_{HWW}^{ice}
         f^{bae}(m_{W_b}^2-m_{W_a}^2-m_{W_e}^2)\\
    + \frac{1}{2m_{W_e}^2}g^{ieb}_{HWW}f^{cae}(m_{W_c}^2-m_{W_a}^2-m_{W_e}^2)\\
    - g_{HWW}^{jab} T^c_{ji}+T_{ji}^bg_{H WW}^{jac}=-f^{bce}g_{H WW}^{iae}
  \end{aligned}
\end{equation}
The $ab$ component results from the the $4W$ WI with 3 contractions given
in the main text in~\eqref{eq:gold-lie}.
\subsection{Scalar potential}
The components of the invariance condition~\eqref{eq:quartic-scalar}
of the scalar potential are obtained by the WIs with external Higgs
bosons and the WI with 4 contractions:
\begin{equation}
  \parbox{20mm}{%
    \begin{fmfchar*}(20,20)
      \fmfleft{A1,A2}
      \fmfright{A3,A4}
      \fmf{double}{A1,a}
      \fmf{dashes,label=$H_i$,label.side=right}{A2,a}
      \fmf{dashes,label=$H_k$}{A3,a}
      \fmf{dashes,label=$H_j$,label.side=left}{A4,a}
      \fmfv{decor.shape=circle,decor.filled=empty,decor.size=25}{a}
    \end{fmfchar*}}\qquad
  \parbox{20mm}{%
    \begin{fmfchar*}(20,20)
      \fmfleft{A1,A2}
      \fmfright{A3,A4}
      \fmf{double}{A1,a}
      \fmf{dashes,label=$H_i$,label.side=right}{A2,a}
      \fmf{double}{A3,a}
      \fmf{dashes,label=$H_j$,label.side=left}{A4,a}
      \fmfv{decor.shape=circle,decor.filled=empty,decor.size=25}{a}
    \end{fmfchar*}}\qquad
  \parbox{20mm}{%
    \begin{fmfchar*}(20,20)
      \fmfleft{A1,A2}
      \fmfright{A3,A4}
      \fmf{double}{A1,a}
      \fmf{dashes,label=$H_i$,label.side=right}{A2,a}
      \fmf{double}{A3,a}
      \fmf{double}{A4,a}
      \fmfv{decor.shape=circle,decor.filled=empty,decor.size=25}{a}
    \end{fmfchar*}}\qquad
  \parbox{20mm}{%
    \begin{fmfchar}(20,20)
      \fmfleft{A1,A2}
      \fmfright{A3,A4}
      \fmf{double}{A1,a}
      \fmf{double}{A2,a}
      \fmf{double}{A3,a}
      \fmf{double}{A4,a}
      \fmfv{decor.shape=circle,decor.filled=empty,decor.size=25}{a}
    \end{fmfchar}}
\end{equation}
Since the components of~\eqref{eq:quartic-scalar} written in terms of
the input parameters~\eqref{eq:input} are somewhat involved and not
needed in this work we refer the reader to~\cite{Schwinn:2003} for the
explicit expressions.


\end{fmffile}
\end{document}